\documentclass[10pt,journal,compsoc]{IEEEtran}

\usepackage{setspace,cite}

\usepackage[table]{xcolor}

\usepackage{pdfpages}
\usepackage{enumitem}
\usepackage{bbm}
\usepackage{url}

\usepackage[tbtags]{amsmath}
\usepackage{amsbsy}
\usepackage{amssymb}
\usepackage{amsfonts}
\usepackage{amsthm}

\usepackage{graphicx}
\usepackage{algorithmic}
\usepackage{algorithm}
\usepackage{balance}
\usepackage{rotating}
\usepackage{multirow}
\usepackage{subcaption}
\usepackage{fancyvrb}
\usepackage{latexsym}
\usepackage{verbatim}
\usepackage{listings}

\newtheorem{lemma}{Lemma}

{\begin{list}               
    {$\bullet$ \hfill}{
        \setlength{\leftmargin}{\parindent}
        \setlength{\parsep}{0.04\baselineskip}
        \setlength{\itemsep}{0.5\parsep}
        \setlength{\labelwidth}{\leftmargin}
        \setlength{\labelsep}{0em}}
    }
{\end{list}}

\providecommand{\eref}[1]{\eqref{#1}}  
\providecommand{\cref}[1]{Chapter~\ref{#1}}

\providecommand{\fref}[1]{Figure~\ref{#1}}

\providecommand{\E}{\ensuremath{\mathbb{E}}}

\providecommand{\bydef}{\overset{\text{def}}{=}}

\renewcommand{\vec}[1]{\ensuremath{\boldsymbol{#1}}}
\providecommand{\mat}[1]{\ensuremath{\boldsymbol{#1}}}

\providecommand{\calD}{\mathcal{D}}

\providecommand{\calF}{\mathcal{F}}

\providecommand{\calN}{\mathcal{N}}

\providecommand{\mC}{\mat{C}}

\providecommand{\mI}{\mat{I}}

\providecommand{\mR}{\mat{R}}

\providecommand{\va}{\vec{a}}
\providecommand{\vb}{\vec{b}}

\providecommand{\vf}{\vec{f}}

\providecommand{\vr}{\vec{r}}

\providecommand{\vv}{\vec{v}}

\providecommand{\vx}{\vec{x}}

\providecommand{\valpha}{\vec{\alpha}}

\providecommand{\vtheta}{\vec{\theta}}

\providecommand{\vxi}{\vec{\xi}}

\providecommand{\vrho}{\vec{\rho}}
\providecommand{\vvarrho}{\vec{\varrho}}

\providecommand{\vzero}{\vec{0}}

\usepackage{array}
\newcommand{\PreserveBackslash}[1]{\let\temp=\\#1\let\\=\temp}
\newcolumntype{C}[1]{>{\PreserveBackslash\centering}p{#1}}
\newcolumntype{R}[1]{>{\PreserveBackslash\raggedleft}p{#1}}
\newcolumntype{L}[1]{>{\PreserveBackslash\raggedright}p{#1}}

\begin{document}

\title{Simulating Anisoplanatic Turbulence by  Sampling Inter-modal and Spatially Correlated Zernike Coefficients}
\author{Nicholas Chimitt,~\IEEEmembership{Student Member,~IEEE}, and Stanley H. Chan,~\IEEEmembership{Senior Member,~IEEE}
\thanks{The authors are with the School of Electrical and Computer Engineering, Purdue University, West Lafayette, IN 47907, USA. Email: \texttt{\{ nchimitt, stanchan\}@purdue.edu}. This work is supported, in part, by the National Science Foundation under grants CCF-1763896 and CCF-1718007.}
}
\graphicspath{{./pix/}}

\IEEEtitleabstractindextext{\begin{abstract}
Simulating atmospheric turbulence is an essential task for evaluating turbulence mitigation algorithms and training learning-based methods. Advanced numerical simulators for atmospheric turbulence are available, but they require evaluating wave propagation which is computationally expensive. In this paper, we present a propagation-free method for simulating imaging through turbulence. The key idea behind our work is a new method to draw inter-modal and spatially correlated Zernike coefficients. By establishing the equivalence between the angle-of-arrival correlation by Basu, McCrae and Fiorino (2015) and the multi-aperture correlation by Chanan (1992), we show that the Zernike coefficients can be drawn according to a covariance matrix defining the correlations. We propose fast and scalable sampling strategies to draw these samples. The new method allows us to compress the wave propagation problem into a sampling problem, hence making the new simulator significantly faster than existing ones. Experimental results show that the simulator has an excellent match with the theory and real turbulence data.
\end{abstract}

\begin{IEEEkeywords}
Atmospheric turbulence, simulator, anisoplanatism, Zernike polynomials, spatially varying blur
\end{IEEEkeywords}}

\maketitle
\section{Introduction}
\subsection{Motivation and Contributions}
Atmospheric turbulence is one of the major challenges for long-range imaging applications in astronomy, surveillance, and navigation. Methods for mitigating atmospheric turbulence have been studied for decades, including many new image processing algorithms \cite{Milanfar2013,Lou2013,Anantrasirichai2013,Lau2017}. Evaluating these algorithms requires ground truth data, and a numerical simulator which can simulate the turbulence from these ground truth data. A simulator is also useful for training supervised learning methods such as deep neural networks because it allows us to synthesize training samples. However, despite a number of turbulence simulators reported in the literature \cite{Potvin_Forand_Dion_2011, Leonard_Howe_Oxford, Repasi_Weiss, Bos_Roggemann, HardieSimulator, Schwarzman2017ICCP, Lachinova2017, Wavetrain, Atcom}, their objectives are usually aimed at providing the best general descriptions of the physical situations. To a large extent, these accurate simulators are overkill for developing image restoration algorithms where the precision may not be observable after the numerical image reconstruction steps. Meanwhile, simple rendering techniques used in computer graphics to synthesize visually pleasing images \cite{Milanfar2013,Lou2013,Anantrasirichai2013,Lau2017} fail to encapsulate the anisoplanatic turbulence over a wide field of view which is easily observable in ground-to-ground imaging. Given the lack of open source simulators aimed at the development of image processing algorithms, in particular those learning-based methods, this paper tries to fill the gap by proposing a new simulator that balances accuracy, interpretability, complexity, and speed.

\begin{figure*}[t]
    \centering
    \includegraphics[width=\linewidth]{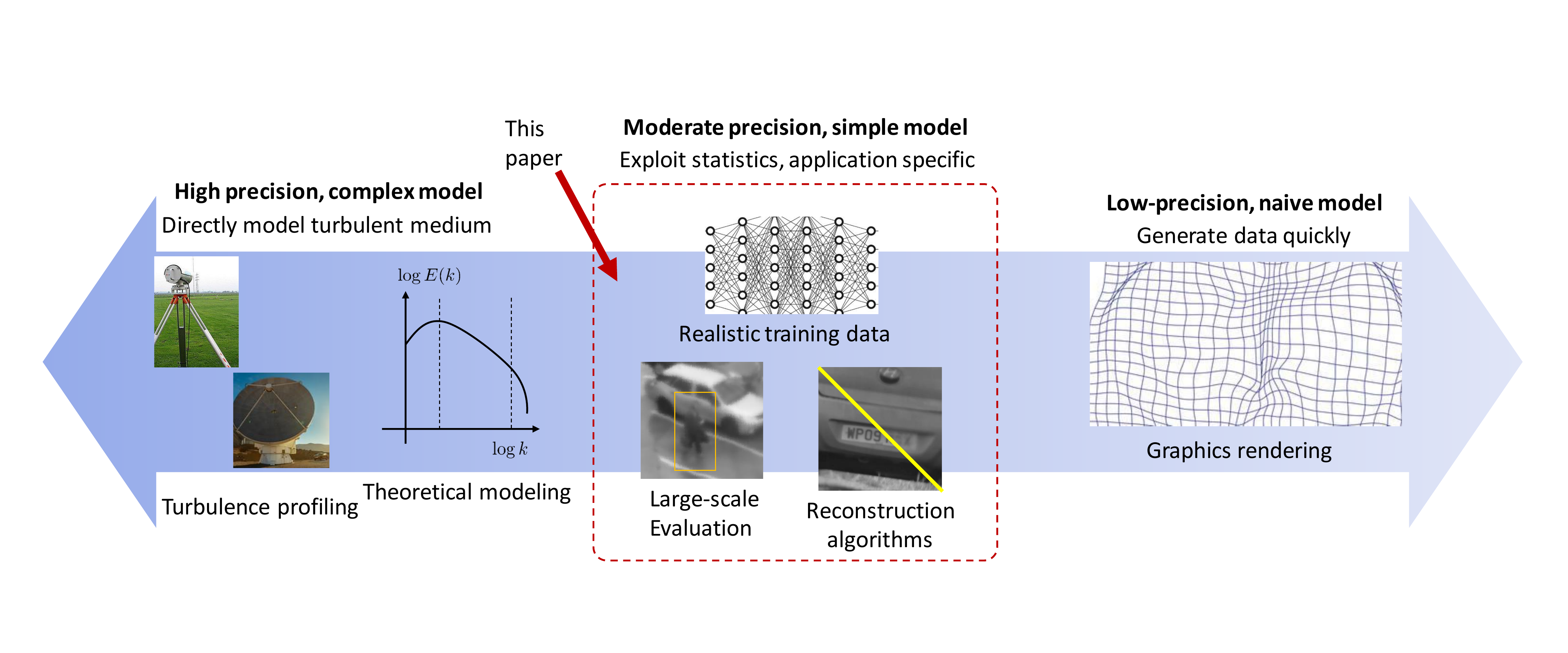}
    \vspace{-8ex}
    \caption{We position this paper in the middle of the spectrum of the literature of turbulence simulators. Our goal is to offer a reasonably accurate turbulence simulator while being fast. In addition, we aim to ensure that our simulator is interpretable so that we can use them to debug image processing algorithms.}
    \label{fig: gtg_ast}
\end{figure*}

Our simulator is based on the idea of directly drawing Zernike coefficients with inter-modal and spatial correlations.  Compared to existing methods, the simulator has two features:
\begin{itemize}
    \item The simulator decouples the tilts and high-order abberations. Based on the classical thin screen model, we leverage a result by Fried (1965) \cite{Fried65} and Noll (1976) \cite{Noll_1976} which showed that the tilts occupy a majority of the energy contained in the turbulent distortions whereas the blurs occupy much less. Because of the significance of the tilts, the decoupling allows us to ``save'' a large number of point spread functions (PSFs) which are originally used to describe the distortions per pixel. In the new simulator, we draw blurs separately from the tilts so that the PSFs are just used to model the high-order aberrations and not ``wasted'' on tilts.
    \item The simulator uses a new approach to ensure the spatial correlations of the tilts. While many existing methods draw tilts by assuming spatial independence (for convenience), we manage to model the spatial correlation without compromising the speed. Our idea is to establish a new theoretical result which connects the work of Basu, McCrae and Fiorino (2015) \cite{Basu_2015} and the work of Chanan (1992) \cite{chanan92}. Furthermore, by using Takato and Yamaguchi (1995) \cite{Takato1995} we show that the spatial correlations can be extended from tilts (the 2nd and the 3rd Zernike coefficients) to \emph{all} high-order Zernike coefficients. Based on these theoretical findings, we develop an algorithmic method to construct the covariance matrices and fast sampling techniques to draw samples from these covariance matrices.
\end{itemize}

The literature of atmospheric turbulence is very rich. However, this has also created a steep learning curve for researchers coming from a non-optics background. One of the motivations of this paper is to promote the subject by making the concepts of atmospheric turbulence transparent and accessible. As such, we devote Section 2 of the paper as a tutorial to help beginning readers. In addition, we provide detailed MATLAB implementations of each of the key steps so that the results are reproducible. We hope that the broad optics community and the image processing community can benefit from this work.

This work is largely aimed towards ground-to-ground imaging, where the medium may be assumed to be of the same strength along the path of propagation. We assume that the depth of the objects is small relative to optical path. Generalization to other problems such as astronomical imaging is important but is not considered in this paper. Moreover, this paper mainly focuses on the spatial correlations. We leave the temporal correlations as a future work. Finally, our results are derived for incoherent imaging as it is common to most of passive imaging systems.

\subsection{Related Work and Position of Our Simulator}
The atmosphere is a turbulent, inhomogeneous medium in which the indices of refraction change spatio-temporally as a function of temperature, pressure, humidity, and wind velocity. As a wave propagates through the turbulent atmosphere, the phase will experience spatial and temporal distortions due to the randomly distributed refractive indices. In order to model the propagation of the wave computationally, one of the most systematic approaches is to split the propagation path into segments and simulate the phase distortion for each segment. Such an approach is known as the split-step propagation. Because of the stage-wise simulation of the propagation path, the split-step propagation is able to capture the turbulent phenomenon to a great extent.

There are plenty of prior works on split-step propagation, including the book by Schmidt
\cite[Chapter 9]{SchmidtTurbBook}, and the work of Bos and Roggemann \cite{Bos_Roggemann}. Since split-step propagation aims to model how a wave propagates through the turbulence, many properties of the turbulence such as spatial and temporal correlations can be captured. However, the biggest challenge of split-step propagation is the computational cost. Split-step propagation involves taking forward and inverse Fourier transforms back and forth multiple times in order to simulate the point spread function (PSF). Since the PSF formation is per pixel, the number of PSFs scales with the number of pixels in an image. As reported by Hardie et al. \cite{HardieSimulator}, the latest split-step simulator takes approximately 120 seconds for a $256 \times 256$ image on a desktop computer.

The alternative approaches to the split-step propagation are propagation-free simulators. However, most of the available propagation-free methods are isoplanatic, i.e. spatially invariant. For example, the simulator developed by the Fraunhofer Research Institute for Optronics and Pattern Recognition \cite{Repasi_Weiss} (and improvements \cite{Leonard_Howe_Oxford}) draws independent and identically distributed (i.i.d.) tilts and simulates the blur using spatially invarying PSFs. The problem is that real turbulent tilts and their blurs are spatially correlated, thus the i.i.d. assumption severely limits the practicality. The method by Potvin et al. \cite{Potvin_Forand_Dion_2011} models the PSF as a bivariate Gaussian with random mean and covariance. For weak turbulence, the bivariate Gaussian can model reasonably well. However, as the turbulence complexity increases, a simple bivariate Gaussian will become inadequate. Some computer vision methods, e.g., \cite{Milanfar2013, Lau2017}, model the tilt using non-rigid deformations. These methods, however, do not closely adhere to the physical turbulence model.

\begin{figure*}[t]
    \centering
    \includegraphics[width = \linewidth]{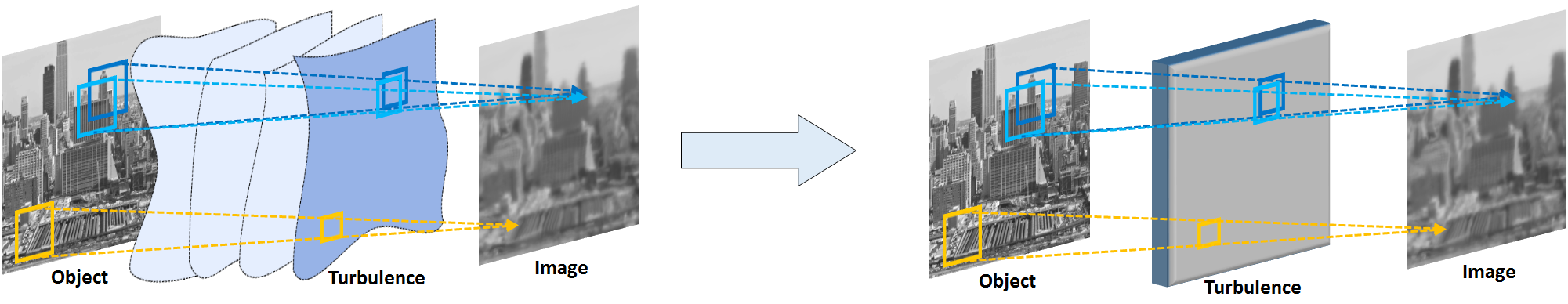} \\
    \caption{[Left] Visualization of how turbulent images are formed. Nearby pixels (the blue boxes) share similar distortions, whereas the far away pixel (the yellow box) has less spatial correlation with the blue pixels. [Right] The thin screen model suggests that the turbulence medium can be modeled as a thin screen. Thus, by analyzing the phase distortions caused by the thin screen we can approximately model the turbulent effects.}
    \label{fig:turbulence_how}
\end{figure*}

Beyond these mainstream simulators, there are other approaches. The work by Schwartzman et al. \cite{Schwarzman2017ICCP} draws spatially correlated tilts by indirectly estimating the angle-of-arrival correlations. However, there is limited quantitative verification besides a few visual comparisons. The brightness function method by Lachinova et al. \cite{Lachinova2017} traces the rays by solving differential equations of the associated waves. Potvin \cite{Potvin_2020} proposed using a first-order approximation of the Born expansion of paraxial propagation.
The simulator proposed by Hunt et al. \cite{Hunt_Iler_SparseRepresntation} statistically learns the sparse representation of the spatially varying point spread functions. However, the training requires a lot of samples and little insight is gained through the approach.

\fref{fig: gtg_ast} illustrates the position of this paper among the other works in the literature. Along the spectrum of methods, we acknowledge the high-precision and complex models offered by split-step and ray tracing type of simulators. These methods are very valuable in computing turbulent conditions in the most general situations. Towards the other end of the spectrum we have the graphics / image processing techniques where the goal is to render a visually appealing image. These simulators are fast and creates great visual effects, although the underlying physics of the turbulence is usually omitted. The position of this paper is in the middle of the spectrum. On the one hand, we aim to develop a reasonably accurate model of the turbulence. However, since our goal is not to provide the best prediction of the turbulence, it becomes unnecessary to compete directly with the those high-precision models. On the other hand, we like the simulator to be as fast as possible so that we can generate synthetic samples for learning-based algorithms. In addition, we aim to ensure that our simulator is easily interpretable so that anyone with limited optics background can understand.

\section{How Waves Propagate through Turbulence: Basic Concepts}
The objective of this section is to provide a brief tutorial of the image formation process when taking an image through turbulence. We provide specific equations of the references so that readers can directly approach the background material. Much of the details can be found in textbooks such as Roggemann's book \cite[Chapters 2,3]{roggemann1996imaging} and Goodman's book \cite[Chapters 4,5,8]{Goodman_StatisticalOptics}.

There are two important key ideas of imaging through turbulence. The first being the correlation between pixels in an image. The second is visualizing the phase distortions. As illustrated in \fref{fig:turbulence_how}, two nearby blocks of the pixels in the object plane will suffer from similar distortions, whereas farther away the distortions are typically independent. If a wave were to be propagating through a vacuum it would arrive at our imaging device roughly flat, though the atmosphere ``dents" this flat wavefront, causing distortions in the image formation process. To describe the turbulent effect, a classical approach is to collapse the turbulent medium into a \emph{thin screen}.
The phase of the thin screen will cause random shifts or ``tilts" and blurs of the pixels.

\subsection{Notations}
We use the following set of notations. We denote the spatial coordinates as $\vx = [x_1, x_2]^T$, and spatial frequencies as $\vf = [f_1, f_2]^T$. The $\ell_2$-norms (i.e., the magnitudes) of the coordinates are denoted as $|\vx|$ and $|\vf|$, respectively. Polar coordinates are denoted using Greek letters $\vxi = [s, \psi]^T$ such that $x_1 = s\cos\psi$ and $x_2 = s\sin\psi$, or $\vvarrho = [r, \theta]^T$ such that $f_1 = r\cos\theta$ and $f_2 = r\sin\theta$. For any scalar field $u(\vx)$, its 2D Fourier transform is
\begin{equation}
    U(\vf) \bydef \calF\left\{u(\vx)\right\} =  \int_{-\infty}^{\infty} u(\vx) e^{-2\pi i \vf^T \vx} d\vx.
\end{equation}
If the scalar field $u(\vx)$ is spherically symmetric, then $u(\vx) = u(|\vx|)$ for all $\vx$.

To model the optical system, we use $D$ to denote the aperture diameter, and $d$ to denote its focal length. The wavelength is $\lambda$, and the wave number is $k \bydef 2\pi / \lambda$. If the optical system has a circular aperture, then the pupil function at the aperture is defined as
\begin{equation}
    W(\vf) = \mbox{circ}\left(\frac{ |\vf|}{D/2}\right)
    = \begin{cases}
    1, \quad\mbox{if}\quad |\vf| \le D/2,\\
    0, \quad\mbox{if}\quad |\vf| > D/2.
    \end{cases}
\end{equation}
The inverse Fourier transform of $W(\vf)$ is \cite[Eq. 2.31]{roggemann1996imaging}
\begin{equation}
    w\left(\vx\right) = \frac{D^2}{4} J_1\left(2\pi \frac{|\vx|D}{2}\right) \Big/ \left(\frac{|\vx|D}{2}\right),
    \label{eq: bessel}
\end{equation}
where $J_1(\cdot)$ is the Bessel function of the first kind.

\subsection{Optical Transfer Functions (OTF)}
As a plane wave propagates through the atmosphere to an observer, it will accumulate a random phase distortion determined by the random phase screens \cite[Section 8.3]{Goodman_StatisticalOptics}.  Multiplying the phase with the pupil function, the wavefront seen by the observer is in the form of
\begin{equation}
    U(\vf) = W(\vf)e^{i \phi(\vf)},
    \label{eq:U(f)}
\end{equation}
where $\phi(\vf)$ denotes the phase distortion function. Assuming incoherent illumination, the observed wavefront can be analyzed (with narrowband assumption) through the optical transfer function (OTF), defined as \cite[Eq. 8.1-3]{Goodman_StatisticalOptics}
\begin{equation}
    H(\vf)
    = \frac{\int U(\vv)U^*(\vv-\lambda d \vf) d\vv}{\int |U(\vv)|^2 d\vv} ,
    \label{eq:OTF 1}
\end{equation}
with $\lambda$ as the mean value of the spectrum of wavelengths.
Substituting \eref{eq:U(f)} into \eref{eq:OTF 1}, and recognizing that \eref{eq:OTF 1} is a convolution, we can write the OTF as \cite[Eq. 2.44]{roggemann1996imaging}
\begin{equation}
    H(\vf) = \frac{W(\lambda d \vf)e^{i\phi(\lambda d \vf)} \ast W(\lambda d \vf)e^{i\phi(\lambda d \vf)}}{W(0)e^{i\phi(0)} \ast W(0)e^{i\phi(0)} }.
    \label{eq: OTF 2}
\end{equation}
The inverse Fourier transform of the OTF is the point spread function (PSF), defined as $h(\vx) = \calF^{-1} \left[ H(\vf) \right]$.

Note that in atmospheric turbulence the phase distortion function $\phi(\vf)$ is a random process. Therefore, the obtained OTF and PSF are also random. To differentiate these random realizations and their ensemble averages, we refer to the former as the \emph{instantaneous} OTF (or PSF), and the latter as the \emph{average} OTF (or PSF).

\vspace{2ex}
\noindent\textbf{Example 1}. As an example of how the phase distortion function can affect the observed image, we consider the special case where $\phi(\vf) = -\frac{2\pi}{\lambda d}\valpha^T \vf$ for some random vector $\valpha$. Substituting this $\phi(\vf)$ into \eref{eq:U(f)} will give us
\begin{equation*}
U(\vf) = W(\vf)e^{i \phi(\vf)} = W(\vf)e^{-i \frac{2\pi}{\lambda d} \valpha^T\vf},
\end{equation*}
where the inverse Fourier transform of $U(\lambda d \vf)$ is
\begin{equation}
\calF^{-1}\left[W(\lambda d \vf)e^{-i2\pi \valpha^T\vf} \right] = w\left(\frac{\vx-\valpha}{\lambda d}\right).
\end{equation}
Now, for simplicity let us assume that $D = 2$ so that $w(\vx)$ is simplified to $w(\vx) = J_1(2\pi|\vx|)/|\vx|$. It follows that the instantaneous PSF is
\begin{equation}
h(\vx) = J_1\left(2\pi \frac{|\vx-\valpha|}{\lambda d}\right)^2 \Big/ \left(\frac{|\vx-\valpha|}{\lambda d}\right)^2.
\label{eq: example PSF}
\end{equation}
Therefore, $\phi(\vf) = -\frac{2\pi}{\lambda d}\valpha^T \vf$ introduces a random tilt to the PSF. The amount of the random tilt is specified by $\valpha$. \hfill $\square$

The instantaneous OTF consists of two parts: (i) The atmospheric OTF $H_{\mathrm{atm}}^{\phi}(\vf)$ due to the turbulence, and (ii) Diffraction limited OTF $H_{\mathrm{dif}}(\vf)$ due to the finite aperture of the optical system. Mathematically, we can write $H(\vf)$ as
\begin{equation}
H(\vf) = H_{\mathrm{atm}}^{\phi}(\vf) H_{\mathrm{dif}}(\vf).
\end{equation}
We will discuss $H_{\mathrm{atm}}^{\phi}(\vf)$ in the next subsection. The diffraction limited OTF $H_{\mathrm{dif}}(\vf)$ is defined as the OTF when there is no turbulence, i.e., setting $\phi(\vf) = 0$. Substituting $\phi(\vf) = 0$ into \eref{eq: OTF 2} and by using \cite[Eq. 3.17]{Fried66optical} we have
\begin{align}
H_{\mathrm{dif}}(\vf)
&\bydef \frac{ W(\lambda d \vf) \ast W(\lambda d \vf) }{ W(0) \ast W(0) } \label{eq: H dif}\\
&= \frac{2}{\pi}\left[\cos^{-1}\left(\frac{|\vf|}{f_c} - \frac{|\vf|}{f_c}\sqrt{1-\left(\frac{|\vf|}{f_c}\right)^2}\right)\right], \notag
\end{align}
for $|\vf| \le f_c$ and is zero otherwise. The cutoff frequency is defined as $f_c = D/(\lambda d)$.

\vspace{2ex}
\noindent\textbf{Example 2}. As a sanity check, we can let $\valpha = \vzero$ in Example 1 so that $\phi(\vf) = 0$. Then \eref{eq: example PSF} reduces to a modified Bessel function in the form of \eref{eq: bessel}. Taking the Fourier transform of $h(\vx)$, with the aid of \cite[Eq. 2.50]{roggemann1996imaging}, will yield an expression exactly equal to $H_{\mathrm{dif}}(\vf)$ in \eref{eq: H dif}. \hfill $\square$

\subsection{Structure Function}
We now discuss $H_{\mathrm{atm}}^{\phi}(\vf)$. The statistics of $H_{\mathrm{atm}}^{\phi}(\vf)$ is detailed in the classic book by Tatarski (1967) \cite{Tatarskii1967}. In particular, Tatarski argued that the random phase distortion function $\phi(\vf)$ is a Gaussian process. Furthermore, for analytic tractability it is often assumed that $\phi(\vf)$ is homogeneous (i.e., wide-sense stationary) and is isotropic (i.e., spherically symmetric). Under these assumptions, the statistics of $\phi(\vf)$ can be fully characterized by its structure function \cite[Eq. 1.13]{Tatarskii1967}
\begin{equation}
    \calD_{\phi}(\vf, \vf') = \E[(\phi(\vf)-\phi(\vf'))^2],
\end{equation}
or simply as $\calD_{\phi}(\vf - \vf')$ if $\calD_{\phi}$ is homogeneous, and $\calD_{\phi}(|\vf - \vf'|)$ if $\calD_{\phi}$ is both homogeneous and isotropic. The structure function has a subtle difference compared to the autocorrelation function. In turbulence,  the structure function is preferred as it is more convenient for processes with stationary increments \cite[p.9]{Tatarskii1967}.

By adopting the Kolmogorov spectrum (1941) \cite{Kolmogorov1941}, Fried showed in his 1966 paper that the structure function can be expressed as \cite[Eq. 5.3]{Fried66optical}
\begin{equation}
    \calD_{\phi}(|\vf - \vf'|) = 6.88(|\vf - \vf'|/r_0)^{5/3},
    \label{eq: structure function}
\end{equation}
where $r_0$ is the atmospheric coherence diameter (or simply the Fried parameter). The Fried parameter is a measure of the quality of optical transmission through turbulence. A small $r_0$ relative to the aperture indicates a high level of blur and tilt variance. The Fried parameter is \cite[Eq. 3.67]{roggemann1996imaging}
\begin{equation}
r_0 = \left[0.423k^2 \int_0^L C_n^2(z) \left(\frac{z}{L}\right)^{5/3} dz\right]^{-3/5}.
\end{equation}
In this equation, $L$ is length of the propagation path, and $C_n^2(z)$ is the refractive index structure parameter. For horizontal imaging, $C_n^2$ can be assumed a constant within a typical acquisition period (e.g., minutes). A typical range of $C_n^2$ is from $1 \times 10^{-14}$m$^{-2/3}$ for strong turbulence to $1 \times 10^{-17}$m$^{-2/3}$ for weak turbulence \cite{Bos_Roggemann}.

\vspace{2ex}
\noindent\textbf{Example 3}. In MATLAB, the Fried parameter can be determined using the command \texttt{integral}. If we let $C_n^2(z) = 1 \times 10^{-15}\mbox{m}^{-2/3}$ for all $z$, $\lambda = 525$nm, and $L = 7000$m, then the Fried parameter is $r_0 = 0.0478$m.  For an aperture of $D = 0.2034$m, the ratio $D/r_0$ is approximately $D/r_0 = 4.26$. This means the smallest spot (i.e., the diffraction PSF) increases 4.26 times due to the turbulence. \hfill $\square$

\begin{figure*}[t]
    \centering
    \includegraphics[width = \linewidth]{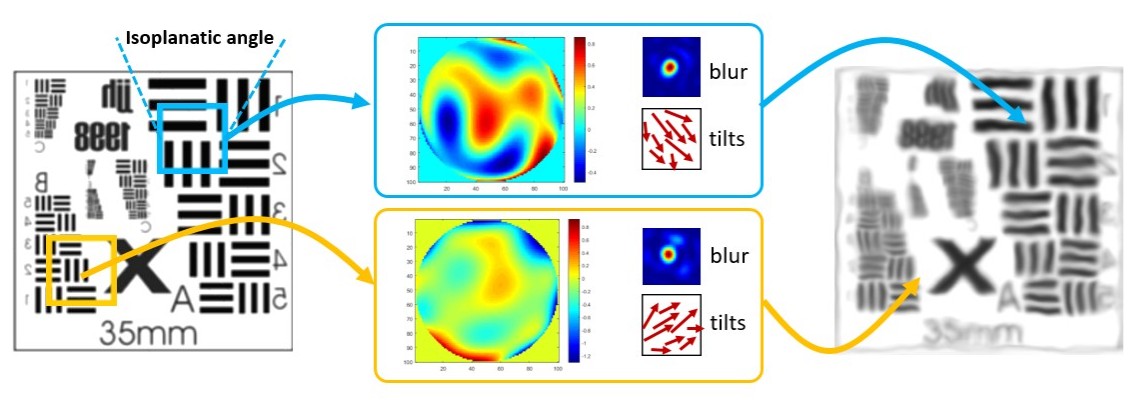} \\
    \caption{The proposed simulator draws random phase distortions according to the distributions of the Zernike coefficients. To speed up the sampling process, the tilts and the tilt-free blurs are drawn separately. The tilts will have spatial correlations, whereas the blurs will be shared within the isoplanatic angle.}
    \label{fig:overview_02}
\end{figure*}

\subsection{Long and Short Exposure OTFs}
The instantaneous atmospheric OTF has a long-term average.
Using the structure function $D_{\phi}(|\vf-\vf'|)$ defined in \eref{eq: structure function}, Fried showed that the \emph{average} atmospheric OTF, which is also called the long-exposure OTF, is \cite[Eq. 3.16]{Fried66optical}
\begin{equation}
\hspace{-2ex} H_{\mathrm{LE}}(\vf) \bydef \E\left[ H_{\mathrm{atm}}^{\phi}(\vf) \right] = \exp\left\{-3.44\left(\frac{\lambda d |\vf|}{r_0}\right)^{5/3}\right\},
\label{eq: HLE}
\end{equation}
where the expectation is taken over the random phase distortion $\phi(\vf)$. Note that $H_{\mathrm{LE}}(\vf)$ is spatially invariant.

If we remove the tilt from the phase of the observed wavefront by defining
\begin{equation}
\varphi(\vf) \bydef \phi(\vf) - \valpha^T\vf,
\end{equation}
where $\valpha^T\vf$ is the best linear fit to $\phi(\vf)$, then the resulting instantaneous atmospheric OTF will not cause any pixel displacement. Taking the expectation over $\varphi(\vf)$ will yield what is referred to as the short-exposure OTF, first proved by Fried in 1966 \cite[Eq. 5.9a]{Fried66optical}:
\begin{align}
&H_{\mathrm{SE}}(\vf)
\bydef \E\left[ H_{\mathrm{atm}}^{\varphi}(\vf) \right] \notag \\
&= \exp\left\{-3.44\left(\frac{\lambda d |\vf|}{r_0}\right)^{5/3} \left[1-\left(\frac{\lambda d |\vf|}{D}\right)^{1/3}\right] \right\} \label{eq: HSE}.
\end{align}

In summary, we note that the turbulent effects are driven by two factors. The first factor is the random tilts which cause pixels to jitter. The average of these jittered PSFs leads to the long-exposure PSF. The other being the high-order abberations which cause the PSF to have random shapes. Its average is the short-exposure PSF.

\subsection{Overview of the Proposed Simulator}
In order to generate the desired distortions at each pixel, one needs to draw the random phase distortion and correspondingly create the point spread function. However, doing so requires running Fourier transforms between the phase domain and the spatial domain for each pixel which is computationally expensive. The proposed simulator alleviates the difficulty by analyzing the Zernike coefficients. Using the classic result of Noll \cite{Noll_1976}, we observe that the 2nd and the 3rd Zernike coefficients are responsible for the tilt, whereas other high-order Zernike coefficients are responsible for the blur. By decoupling the tilts and the blurs, we are able to speed up the simulator. \fref{fig:overview_02} provides a pictorial illustration.

\begin{itemize}
    \item \textbf{The 2nd and 3rd Zernike Coefficients}. The tilts (the 2nd and the 3rd Zernike coefficients) caused by turbulence are spatially correlated throughout the entire image. To this end, we develop a sampling scheme which allows us to compute the spatial correlation matrix and consequently draw samples. However, constructing the spatial correlation matrix requires analyzing the statistics of the angle of arrivals. We showed how the angle of arrivals can be linked to the multi-aperture concept. Consequently, we showed that spatial correlations can be extended to all high-order Zernike coefficients.
    \item \textbf{High-order Zernike Coefficients}. For high-order Zernike coefficients, we assume homogeneity within the isoplanatic angle. This is based on the observations made by Noll \cite{Noll_1976}, who showed that majority of the turbulence distortions is occupied by the tilts and not the blur. Thus when tilt is removed, different tilt-free blurs will not have substantially different influence to different pixels within the isolplanatic angle. This is especially true when the digital image has limited resolution.
\end{itemize}

In what follows, we will describe the concept of the Zernike polynomials. Specifically, we will discuss two types of correlations: (i) Intermodal correlation of the Zernike coefficients in Section \ref{sec: zernike 3}, and (ii) Spatial correlation of the Zernike coefficients in Section \ref{sec: tilt 4}.

\section{Sampling Zernike Coefficients with Inter-modal Correlations}
\label{sec: zernike 3}
In this section we present the idea of how to draw Zernike polynomials for simulating turbulence. We will first discuss the original concept proposed by Noll in 1976, with elaborations of the implementations in Section \ref{sec: zernike 3a}. Then, in Section \ref{sec: zernike 3b} we discuss the concept of inter-modal correlations, and discuss how the Zernike coefficients are drawn. The results presented in this section are applicable to all orders of Zernike coefficients.

\subsection{Zernike Polynomials for Phase Distortions}
\label{sec: zernike 3a}
Zernike polynomials are a set of orthogonal functions defined on the unit disk. Because most optical systems have a circular aperture, Zernike polynomials are particularly useful in providing the basis representations.

To work with Zernike polynomials it is more convenient to switch the Cartesian coordinate $\vf = [f_x, f_y]^T$ to the polar coordinate $\vvarrho = [r,\theta]^T$, where $f_x = r\cos\theta$ and $f_y = r\sin\theta$. By further defining $R \bydef D/2$ as the aperture radius, we can scale $r$ to $r = R \rho$ so that $0 \le \rho \le 1$. Therefore, the phase function $\phi(\vf)$ can be written as $\phi(R\vrho)$ where $\vrho = [\rho,\theta]^T$ and $R\vrho = [R\rho,\theta]^T$.

Defining $\{Z_j(\vrho)\}_{j=1}^{N}$ as the Zernike basis and $\{a_j\}_{j=1}^{N}$ as the coefficients, the phase $\phi$ can be represented as
\begin{equation}
    \phi(R\vrho) = \sum_{j=1}^{N} a_j Z_j(\vrho).
\end{equation}
As elaborated by Noll \cite[Table 1]{Noll_1976}, the Zernike coefficients have geometric interpretations. For example, $a_2$ and $a_3$ are the horizontal and vertical tilts. These coefficients can be determined using the orthogonality principle:
\begin{equation}
a_j = \langle \phi, Z_j \rangle_{W}  \bydef \int_{0}^{2\pi}\int_0^1 W(\vrho) \phi(R\vrho) Z_j(\vrho) d\rho d\theta,
\end{equation}
where $\langle \cdot, \cdot \rangle_{W}$ denotes the inner product using the pupil function $W(\cdot)$ as a weight. Since $\phi(R\vrho)$ is zero-mean Gaussian \cite{Tatarskii1967}, the Zernike coefficients (which are obtained through linear projections) are also zero-mean Gaussians.

The following Lemma is useful but seldom mentioned in the literature. We will use it in Section \ref{sec: zernike 3b} and Section \ref{sec: tilt 4}.

\begin{lemma}
The integration of the Zernike polynomials $Z_j(\vrho)$ over the unit disk $W(\vrho)$ is zero for any $j \ge 2$, i.e.,
\begin{equation}
\int W(\vrho) Z_j(\vrho) d\vrho = 0.
\end{equation}
\end{lemma}
\begin{proof}
See Appendix.
\end{proof}

\vspace{2ex}

\begin{figure}[h]
\centering
\begin{tabular}{cc}
\includegraphics[height=3.5cm]{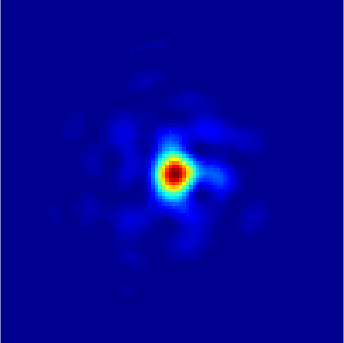}&
\includegraphics[height=3.5cm]{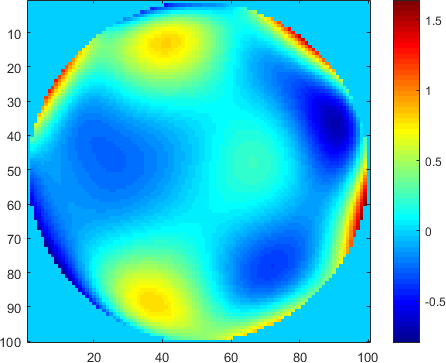}\\
(a) $h(\vx)$ & (b) $\phi(R\vrho)$
\end{tabular}
\caption{The random phase distortion (b) generated from the random Zernike polynomial in a grid of $100 \times 100$ (corresponding to a disk of diameter 2), and its associated PSF (a). See Example 1 for experiment description.}
\label{fig: Zernike and PSF}
\end{figure}

\noindent\textbf{Example 4}. We randomly draw 36 i.i.d. Gaussian coefficients $\{a_j\}_{j=1}^{36}$ from $a_j \sim \calN(0,1/\sqrt{36})$. Then, using the MATLAB code developed by Gray \cite{Gray_MATLAB}, the phase $\phi(R\vrho)$ can be generated using the command \texttt{ZernikeCalc}. Taking the inverse Fourier transform of  $W(\vrho)\exp\{i\phi(R\vrho)\}$ as in \eref{eq: OTF 2}, then its magnitude squared will yield the PSF. \fref{fig: Zernike and PSF} illustrates one random realization of the phase distortion function $\phi(R\vrho)$ and the corresponding PSF $h(\vx)$. The MATLAB code is shown in Algorithm~\ref{alg: phase screen}. Here, we over-sample the Fourier spectrum using \texttt{fftK = 8} for anti-aliasing. The division \texttt{ph/2} is due to the fact that the diameter is 2. \hfill $\square$

\begin{algorithm}[h]
\begin{Verbatim}[fontsize = \footnotesize]
N = 100;
fftK = 8;
[W, ~]  = ZernikeCalc(1,1);
[ph, ~] = ZernikeCalc(1:36, (1/6)*randn(36,1), ...
            N, 'STANDARD');
U   = exp(1i*2*pi*ph/2).*W;
u   = abs(fftshift(ifft2(U, N*fftK, N*fftK).^2));
\end{Verbatim}
\caption{MATLAB code to generate Zernike phase distortion}
\label{alg: phase screen}
\end{algorithm}

While the i.i.d. sampling in Example 4 has demonstrated the usage of the Zernike polynomials, it does not reflect any turbulence physics. In the followings we introduce two types of correlations: (i) Inter-mode correlation: How a coefficient $a_j$ is correlated to another coefficient $a_{j'}$. See Section \ref{sec: zernike 3b}. (ii) Spatial correlation: How a coefficient $a_j(\vr)$ is correlated to itself but at a different location $a_{j}(\vr')$. See Section \ref{sec: tilt 4}.

\subsection{Zernike Inter-Mode Correlations}
\label{sec: zernike 3b}
Inter-mode correlation refers to the correlation between $a_{j}$ and $a_{j'}$ for $j \not= j'$. It can be determined by taking the expectation $\E[a_ja_{j'}]$, as defined by Noll \cite[Eq. 22]{Noll_1976}:
\begin{align}
\E[a_{j}^* a_{j'}]
&= \int \int W(\vrho)W(\vrho') Z_j(\vrho) Z_{j'}(\vrho') \notag \\
&\quad \times \E[\phi(R\vrho)  \phi(R\vrho')] \;\;  d\vrho d\vrho'.
\label{eq: correlation 1}
\end{align}
The expectation inside the double integration of \eref{eq: correlation 1} is the autocorrelation of the phase. The following lemma shows how this autocorrelation can be switched to the structure function. It is an important result, but, like Lemma 1, does not seem to be available in the literature.

\begin{lemma}
The inter-mode correlation $\E[a_{j}^* a_{j'}]$ in \eref{eq: correlation 1} is equivalent to
\begin{align}
\E[a_{j}^* a_{j'}]
&= - \int \int W(\vrho)W(\vrho') Z_j(\vrho) Z_{j'}(\vrho') \notag \\
&\quad \times D_{\phi}(R\vrho-R\vrho') \;\;  d\vrho d\vrho',
\label{eq: correlation 2}
\end{align}
where $D_{\phi}(R\vrho-R\vrho') \bydef \E[(\phi(R\vrho) - \phi(R\vrho'))^2]$.
\end{lemma}
\begin{proof}
See Appendix.
\end{proof}

With Lemma 2, we can substitute the Kolmogorov structure function \eref{eq: structure function} into \eref{eq: correlation 2} to obtain an expression previously derived by Noll \cite[Eq. 25]{Noll_1976} (using Wiener-Khinchin theorem):
\begin{align}
&\E[a_{j}^* a_{j'}] = \left(\frac{0.046}{\pi}\right) \left(\frac{D}{2r_0}\right)^{5/3} [(n+1)(n'+1)]^{1/2} \\
&\times (-1)^{n+n'-2m} \delta_{m,m'} \int_0^{\infty} \zeta^{-8/3} \frac{J_{n+1}(2\pi\zeta) J_{n'+1}(2\pi\zeta)}{\zeta^2} d\zeta, \notag
\end{align}
where $m$, $m'$, $n$, $n'$ are the Azimuthal frequency and radial degree as defined in \cite[Table 1]{Noll_1976}. A closed-form expression of the integral can be obtained using \cite[Eq. A2]{Noll_1976}, of which the MATLAB code is shown below. \fref{fig: inter mode} shows the inter-mode correlation in the log-magnitude scale, i.e., $\log |\E[a_j^* a_{j'}]|$, and the normalized scale, i.e., $\E[a_j^* a_{j'}]/\sqrt{\E[a_j^2]\E[a_{j'}^2]}$ without the factor $(D/r_0)^{5/3}$.

\begin{figure}[h]
\centering
\begin{tabular}{cc}
\includegraphics[height=3.6cm]{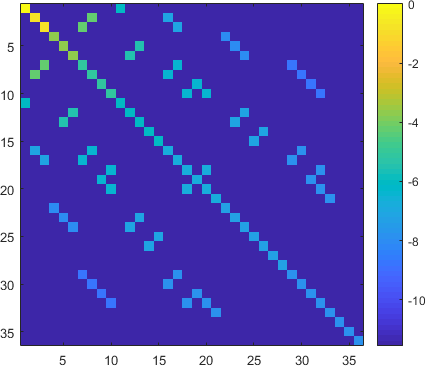}&
\hspace{-2ex}
\includegraphics[height=3.6cm]{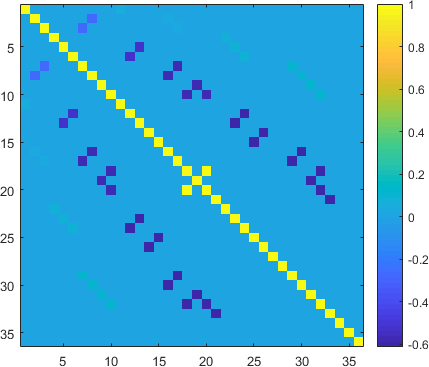}\\
(a)$\log |\E[a_j^* a_{j'}]|$ & (b) $\E[a_j^* a_{j'}]/\sqrt{\E[a_j^2]\E[a_{j'}^2]}$
\end{tabular}
\caption{Inter-mode correlation of the Zernike coefficients. (a) Log-magnitude plot. (b) Normalized plot.}
\label{fig: inter mode}
\vspace{-2ex}
\end{figure}

\begin{algorithm}[h]
\begin{Verbatim}[fontsize=\footnotesize]
C = zeros(36,36);
for i=1:36, for j=1:36
    ni = n(i); nj = n(j); % n and m are given in
    mi = m(i); mj = m(j); % Tab 1 of [Noll 1976]
    if (mod(i-j,2)~=0)||(mi~=mj)
        C(i,j) = 0;
    else
        den = gamma((ni-nj+17/3)/2)*...
            gamma((nj-ni+17/3)/2)*...
            gamma((ni+nj+23/3)/2);
        num = 0.0072*(-1)^((ni+nj-2*mi)/2)*...
            sqrt((ni+1)*(nj+1))*pi^(8/3)*...
            gamma(14/3)*gamma((ni+nj-5/3)/2);
        C(i,j) = num/den;
    end
end end
\end{Verbatim}
\caption{MATLAB code to generate Zernike Coefficients using \cite[Equation A2]{Noll_1976}}
\end{algorithm}

The collection of the correlations $\E[a_{j}^* a_{j'}]$ defines a covariance matrix $\mC$ where the $(j,j')$-th element is
\begin{equation}
[\mC]_{j,j'} = \E[a_{j}^* a_{j'}].
\label{eq: inter mode corr}
\end{equation}
In order to draw correlated samples from this covariance matrix, we decompose $\mC = \mR\mR^T$ using the Cholesky factorization. Then, for any zero-mean unit-variance Gaussian random vector $\vb \sim \calN(\vzero,\mI)$, the transformed vector
\begin{equation}
\va = \mR\vb
\label{eq: transfomed Gaussian}
\end{equation}
will have the property that $\E[\va\va^T] = \mC$. Therefore, to draw a set of inter-mode correlated Zernike coefficients we simply draw a white noise vector $\vb$ and transform using \eref{eq: transfomed Gaussian}. The computation cost is low because $N$ is usually not large. Algorithm~\ref{alg: random sample} shows the MATLAB implementation for drawing $a_4,\ldots,a_{N}$. These coefficients can be substituted into Algorithm~\ref{alg: phase screen} to generate the phase distortion function.

\begin{algorithm}[h]
\caption{MATLAB code to draw random phase}
\begin{Verbatim}[fontsize=\footnotesize]
R = chol(C(4:36, 4:36));
b = randn(33,1);
a = sqrt((D/r0)^(5/3))*R*b;
[ph, ~] = ZernikeCalc(4:36, a, N, 'STANDARD');
\end{Verbatim}
\label{alg: random sample}
\end{algorithm}

\begin{figure}[h]
\centering
\includegraphics[width=\linewidth]{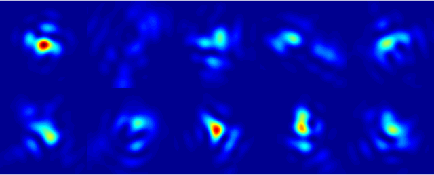}
\caption{Spatially varying PSFs generated by the Zernike coefficients $a_4,\ldots,a_{36}$, using $C_n^2 = 1 \times 10^{-15}$m$^{-2/3}$, $L = 7000$m, $\lambda = 525$nm, $D = 0.2034$m, and $d = 1.2$m.}
\label{fig: PSFs}
\end{figure}

\vspace{2ex}
\noindent\textbf{Example 5}. Using the same configurations as in Example 3, we can draw random PSFs using Algorithm~\ref{alg: random sample}. \fref{fig: PSFs} shows 10 randomly drawn PSFs using Zernike coefficients $a_4,\ldots,a_{36}$. In this particular example, we set the Fourier over-sampling rate \texttt{fftK = 8} for display purpose.

\section{Sampling Zernike Coefficients with Spatial Correlations}
\label{sec: tilt 4}
In Section \ref{sec: zernike 3} we discussed how to draw Zernike coefficients with inter-modal correlations, but just for one pixel. To draw Zernike coefficients for the entire image we need to consider spatial correlations. In this section, we describe a key innovation of this paper, namely, connecting multi-aperture correlations to angle of arrival correlations.

\subsection{Our Plan of Development}
\begin{figure*}[t]
\centering
\includegraphics[width=0.65\linewidth]{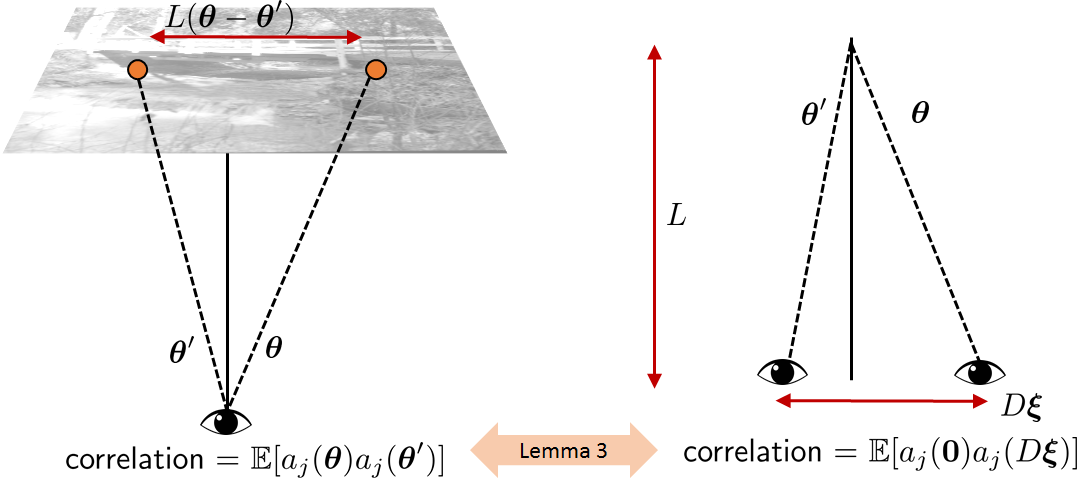}
\vspace{2ex}
\caption{The plan theoretical development: [Left] shows the angle of arrival model, where two points on the object plane travel to one point on the image plane. The separation between the two points are $L(\vtheta-\vtheta')$. [Right] shows the multi-aperture model, where the separation on the image plane is $D\vxi$, where $D$ is the aperture diameter, and $\vxi$ is defined such that $L(\vtheta-\vtheta') = D\vxi$. The correlation in the angle-of-arrival model is what we want, but the existing results are based on the multi-aperture model. Our plan is to establish some form of equivalence between the two correlations. }
\label{fig: aoa concept}
\end{figure*}

Before we discuss the technical details, let us first outline the plan. The spatial correlation problem arises when we consider two points in the object plane as shown in \fref{fig: aoa concept}. Given the angle of arrival angles $\vtheta$ and $\vtheta'$, the separation in the object plane can be computed as $L(\vtheta-\vtheta')$. The corresponding $j$-th Zernike coefficients $a_j(\theta)$ and $a_j(\theta')$ will have a correlation $\E[a_j(\theta) a_j(\theta')]$. However, this correlation is generally unknown except for Zernike coefficients $j = 2$ and $j = 3$. What is known in the literature is the multi-aperture model as shown on the right hand side of \fref{fig: aoa concept}. In this figure, we measure the separation not in the object plane but in the image plane. The separation is measured in terms of the aperture diameter $D$, and the correlation becomes $\E[a_j(\vzero)a_j(D\vxi)]$ where $\vxi$ is chosen such that $L(\vtheta-\vtheta') = D\vxi$. Our goal is to establish some form of equivalence between $\E[a_j(\vtheta)a_j(\vtheta')]$ and $\E[a_j(\vzero)a_j(D\xi)]$.

The approach outlined above is based on a few prior work. The angle of arrival correlation follows from Basu, McCrae and Fiorino (2015) \cite{Basu_2015}, which can be dated back to Fried's original paper on differential angle of arrival in 1976 \cite{Fried1976_variety}. The biggest limitation is that the result is limited to tilts only. While we are mostly interested in the tilts, we believe that a general technique that can be extended to other high-order Zernike coefficients would be useful. To this end, we establish the equivalence with Chanan's multi-aperture approach (1992) \cite{chanan92}. Chanan's result is attractive because it has explicit equations for the spatial correlations of Zernike tilt coefficients. Chanan's result is also extendable to high-order Zernike coefficients according to Takato and Yamaguchi (1995) \cite{Takato1995}.


\subsection{Zernike Correlation of Angle-of-Arrivals}
Consider two points in the object plane with angle-of-arrivals $\vtheta$ and $\vtheta'$, respectively, as shown in \fref{fig: aoa concept}. The distance between the pupil plane and the object plane is $L$.

The spatial correlation of the Zernike coefficients $a_j$ located at $\vtheta$ and $\vtheta'$ is defined as \cite[Eq. 3]{Basu_2015}
\begin{align}
\E[a_j(\vtheta)a_j(\vtheta')] &\bydef \int \int W(\vrho) W(\vrho') Z_j(\vrho) Z_j(\vrho') \notag \\
                            &\quad\quad \times \E[\phi(R\vrho,\vtheta) \phi(R\vrho', \vtheta')] d\vrho d\vrho'.
\end{align}
Adopting the exact argument of Lemma 2, we can show that $\E[a_j(\vtheta)a_j(\vtheta')]$ is equivalent to
\begin{align}
\E[a_j(\vtheta)a_j(\vtheta')] &= -\int \int W(\vrho) W(\vrho') Z_j(\vrho) Z_j(\vrho') \notag \\
                            &\quad\quad \times D_{\phi}(R\vrho-R\vrho', \vtheta-\vtheta') d\vrho d\vrho',
                            \label{eq: Basu Zernike D}
\end{align}
where the new structure function is defined as \cite[Eq. 6]{Basu_2015}
\begin{align}
&D_{\phi}(R\vrho-R\vrho', \vtheta-\vtheta') \bydef 2.91k^2 \int_0^L C_n^2(z) \label{eq: D new}\\
&\quad\quad\quad\quad \times \left| (R\vrho-R\vrho')\left(1-\frac{z}{L}\right) + z(\vtheta-\vtheta')\right|^{5/3} dz. \notag
\end{align}
If we substitute $\vtheta = \vtheta'$, then $D_{\phi}(R\vrho-R\vrho', \vzero)$ is identical to the structure function in \eref{eq: structure function}. We should highlight that \eref{eq: Basu Zernike D} is a new result, as we generalize the tilts in Basu et al. \cite{Basu_2015} to Zernike polynomials $Z_j(\vrho)$ of arbitrary degree.

The next lemma is a major new result. This result allows us to establish the equivalence between the angle-of-arrival result by Basu et al. \cite{Basu_2015} and the multi-aperture result by Chanan \cite{chanan92}. The key idea is to approximate the integration in \eref{eq: D new} so that it takes the form of \eref{eq: structure function} even if $\vtheta \not= \vtheta'$.

\begin{lemma}
Assume a constant $C_n^2$. The structure function in \eref{eq: D new} can be approximated by
\begin{align}
&D_{\phi}(R\vrho-R\vrho', \vtheta-\vtheta') \notag \\
&\quad\quad \cong 2.91k^2 C_n^2 L \left| \frac{(R\vrho-R\vrho')}{2} + \frac{L (\vtheta-\vtheta')}{2} \right|^{5/3}.
\label{eq: D new 2}
\end{align}
\end{lemma}
\begin{proof}
See Appendix.
\end{proof}

\noindent\textbf{Remark}: The approximation error in Lemma 3 is determined by the second order term of the Taylor series, which is upper bounded by $2.91k^2 C_n^2 L |f''(\frac{1}{2})|/24$.

\subsection{Zernike Correlation of Multi-Apertures}
\label{sec: Zernike_multi_ap}

\begin{figure}[t]
\centering
\vspace{-2ex}
\includegraphics[width=0.75\linewidth]{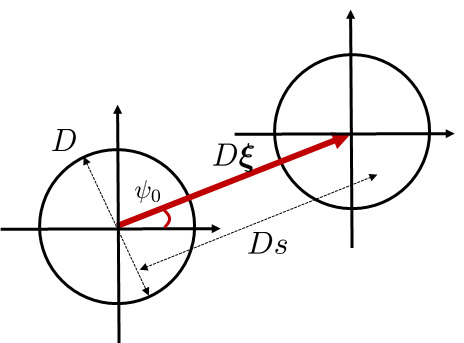}
\caption{Notation of the coordinates. The magnitude of $\vxi$ is measured in the unit of the diameter $D$.}
\label{fig: coordinate}
\vspace{-2ex}
\end{figure}

We are now in a position to transition to Chanan's multi-aperture concept. To start with let us consider the structure function in \eref{eq: D new 2}. Pulling out the factor $2^{5/3}$ the structure function is proportional to $|R\vrho - R\vrho' + L (\vtheta - \vtheta')|^{5/3}$. Defining
\begin{equation*}
D\vxi \bydef L(\vtheta - \vtheta'),
\end{equation*}
where $D$ is the aperture diameter, the structure function in \eref{eq: D new 2} can be written as
\begin{align}
&D_{\phi}(R\vrho-R\vrho' + D\vxi) \notag \\
&\quad\quad\quad = \frac{2.91 k^2 C_n^2 L}{2^{5/3}} \left| R\vrho-R\vrho' + D\vxi \right|^{5/3}.
\end{align}
Without loss of generality we may assume $\vtheta = \vzero$. Thus, the correlation becomes
\begin{align}
\E[a_{j}^* a_{j'}(D\vxi)] &= -\int \int W(\vrho)W(\vrho') Z_j(\vrho) Z_{j'}(\vrho') \notag \\
                          &\quad \times \calD_{\phi}(R\vrho-R\vrho'+D\vxi) \;\;  d\vrho d\vrho',
\label{eq: correlation 2}
\end{align}
which is exactly \cite[Eq. 3]{chanan92} with an additional constant $2^{5/3}$.

Now, following Chanan's derivation \cite[Eq. 11]{chanan92} we show that for $j \in \{2,3\}$ (i.e., the tilts),
\begin{align}
&\E[a_{j}^* a_{j}(D\vxi)] = \frac{c_2}{2^{5/3}} \left(\frac{D}{r_0}\right)^{5/3} \left[I_{0}(s) \mp \cos 2\psi_0 I_2(s)\right],
\label{eq: correlation 3}
\end{align}
where $c_2 = 7.7554$, and the minus sign is for $a_2$ and the plus sign is for $a_3$. The meaning of the coordinate $[s, \psi_0]$ is illustrated in \fref{fig: coordinate}. The length of $D\vxi$ is the scalar $Ds$, and the angle is $\psi_0$. When Chanan first studied the problem he was considering two apertures separated by a vector $D\vxi$.

\begin{figure}[h]
\centering
\includegraphics[width=\linewidth]{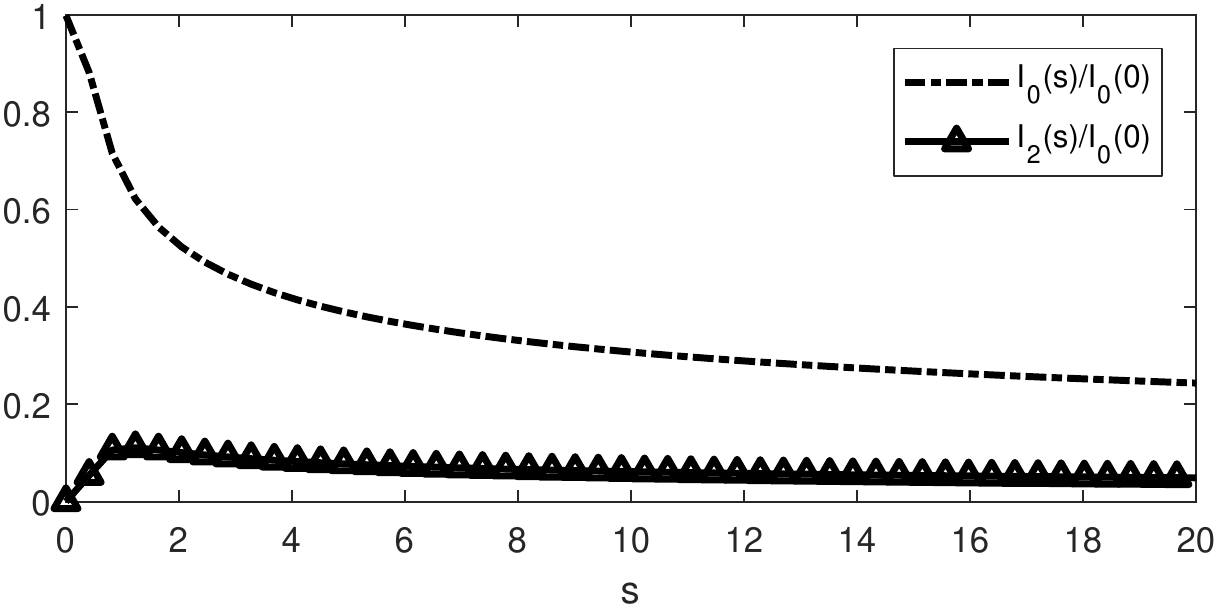}
\caption{The integrals $I_0(s)/I_0(0)$ and $I_2(s)/I_0(0)$.}
\label{fig: I0I2}
\end{figure}

The integrals $I_0(s)$ and $I_2(s)$ in \eref{eq: correlation 3} are defined through the Bessel functions of the first kind:
\begin{align}
I_0(s) &= \int_0^{\infty} \zeta^{-14/3}J_0(2s\zeta)J_2^2(\zeta)d\zeta, \notag\\
I_2(s) &= \int_0^{\infty} \zeta^{-14/3}J_2(2s\zeta)J_2^2(\zeta)d\zeta,
\end{align}
which can be defined in MATLAB using commands \texttt{besselj} and \texttt{integral} as shown in Algorithm~\ref{alg: I0I2}. \fref{fig: I0I2} shows the normalized $I_0(s)/I_0(0)$ and $I_2(s)/I_0(0)$ as a function of $s$.

\begin{algorithm}[h]
\caption{MATLAB code to evaluate integrals $I_0$ and $I_2$}
\begin{Verbatim}[fontsize=\footnotesize]
s   = linspace(0,smax,N);
f   = @(z) z^(-14/3)*besselj(0,2*s*z)*...
                    besselj(2,z)^2;
I0  = integral(f, 1e-8, 1e3, 'ArrayValued', true);
g   = @(z) z^(-14/3)*besselj(2,2*s*z)*...
                    besselj(2,z)^2;
I2  = integral(g, 1e-8, 1e3, 'ArrayValued', true);
\end{Verbatim}
\label{alg: I0I2}
\end{algorithm}

\begin{figure}[t]
\centering
\begin{tabular}{cc}
\includegraphics[height=3.5cm]{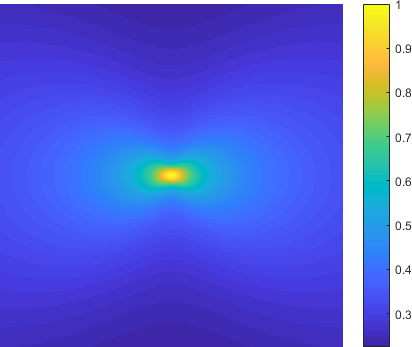}&
\includegraphics[height=3.5cm]{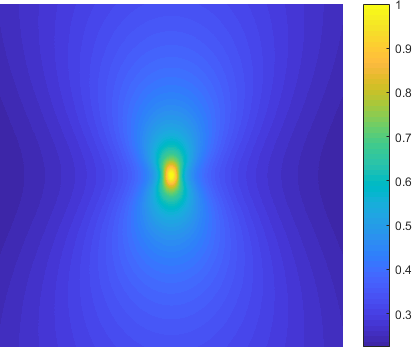}\\
(a) $C_2(\vxi)$ & (b) $C_3(\vxi)$
\end{tabular}
\caption{Spatial correlation of the horizontal and vertical tilts. Shown in this figure are the normalized correlation coefficients (a) $C_2(\vxi)$ and (b) $C_3(\vxi)$.}
\label{fig: Cx and Cy}
\end{figure}

The expression in \eref{eq: correlation 3} provides a convenient way of constructing a covariance matrix. In particular, for $j \in \{2,3\}$, we can consider the normalized correlation coefficients
\begin{equation}
C_j(\vxi) = \frac{\E[a_{j}^* a_{j}(D\vxi)]}{\E[a_{j}^* a_{j}(0)]} = \frac{I_{0}(s) \mp \cos (2\psi_0) I_2(s)}{I_0(0)}.
\end{equation}
Converting the polar coordinate $\vxi$ to Cartesian coordinate we can visualize the normalized correlation coefficient in \fref{fig: Cx and Cy}.
The MATLAB code is shown in Algorithm~\ref{alg: CxCy}.

\vspace{-2ex}

\vspace{2ex}
\begin{algorithm}[h]
\caption{MATLAB code to generate $C_2$ and $C_3$}
\begin{Verbatim}[fontsize = \footnotesize]
i = N/2; j = N/2;
[x,y] = meshgrid(1:N,1:N);
psi   = atan((i-x)./(j-y));
s     = round(sqrt((i-x).^2 + (j-y).^2));
s     = min(max(s,1),N);
C2    = (I0(s)-cos(2*psi).*I2(s))/I0(1);
C2(j,i) = 1;
C3    = (I0(s)+cos(2*psi).*I2(s))/I0(1);
C3(j,i) = 1;
\end{Verbatim}
\label{alg: CxCy}
\end{algorithm}

To draw the random tilts, we need a proper scaling so that the Zernike coefficient can correspond to the number of pixels. As mentioned in \cite[Eq. 2]{chanan92}, the relationship between the tilt angle $\alpha_x$ and the Zernike coefficient is given by
\begin{equation}
\alpha_x \mbox{ [radians]}= \frac{2\lambda}{\pi D}a_2, \quad\mbox{and}\quad \alpha_y = \frac{2 \lambda}{\pi D}a_3.
\end{equation}
The tilt angle $\alpha_x$ and $\alpha_y$ are measured in radians. Since the Nyquist sampling between two adjacent pixels in the object plane is $L\lambda/2D$, it holds that the tilt angles converted to number of pixels in the object are
\begin{align*}
\alpha_x \mbox{ [pixels]} =  \frac{L  \cdot \frac{2\lambda}{\pi D} a_2}{\frac{L\lambda}{2D}} = \frac{4}{\pi}a_2, \quad\mbox{and}\quad \alpha_y \mbox{ [pixels]} = \frac{4}{\pi} a_3.
\end{align*}
Putting everything together, we can show that the un-normalized correlation coefficient of the tilt angles is
\begin{align}
&\E[\alpha_x^* \alpha_x(D\vxi)] = \frac{16}{\pi^2} \E[a^*_2 a_2(D\vxi)] \notag \\
&\quad\quad = \underset{\kappa^2}{\underbrace{\frac{16}{\pi^2} \frac{c_2}{2^{5/3}} I_0(0)}} C_2(\vxi),
\end{align}
where we defined $\kappa$ as the constant preceding the normalized correlation function $C_2(\vxi)$. For vertical tilts, the relationship is $\E[\alpha_y^* \alpha_y(D\vxi)] = \kappa^2 C_3(\vxi)$.

\vspace{2ex}
\noindent\textbf{Remark}: The autocorrelation functions $C_2(\vxi)$ and $C_3(\vxi)$ are anisotropic by construction. This is consistent with Fried's differential angle of arrival result in 1975 \cite[Eq. 32]{Fried1975_differential}. To enforce isotropy one can simply restrict the angle $\psi_0$ to be such that $\vxi$ is always aligned with $\vrho-\vrho'$. In this case we can show that $C_2(\vxi) = C_3(\vxi) = I_0(s) + I_2(s)$.

\subsection{Fast Sampling via Fourier Transforms}
\begin{figure}[h]
\centering
\begin{tabular}{cc}
\includegraphics[height=3.5cm]{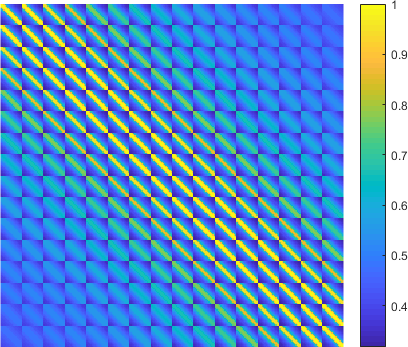}&
\includegraphics[height=3.5cm]{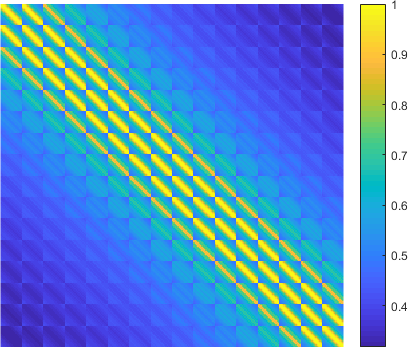}\\
(a) $\mC_2$ &(b) $\mC_3$
\end{tabular}
\caption{The covariance matrix $\mC_2$ and $\mC_3$ for an image of $16 \times 16$ pixels. The size of the matrices are $16^2\times 16^2$.}
\label{fig: SigmaX and SigmaY}
\vspace{-2ex}
\end{figure}

Given the normalized correlation coefficients $C_2(\vxi)$ and $C_3(\vxi)$ we can in principle draw random samples using the same routine outlined in \eref{eq: inter mode corr} and \eref{eq: transfomed Gaussian}. However, complication arises because this would require constructing the covariance matrices $\mC_2$ and $\mC_3$ of sizes $N^2 \times N^2$ where $N$ is the number of pixels of an image. For small problems we can visualize the matrices, e.g., \fref{fig: SigmaX and SigmaY}.

A more scalable approach to draw samples is to recognize that $C_j(\vxi)$ is an autocorrelation function of the Zernike coefficient. By construction, it is wide-sense stationary and so we can draw random samples from its power spectral density. Therefore, given $C_j(\vxi)$ we take the 2D Fourier transform, and multiply the power spectral density with white noise. It is important to note here many other methods for generating the tilts could be considered here, such as sub-harmonic approaches described in \cite[Chapter 9]{SchmidtTurbBook} or more recent approaches such as Paulson et al. \cite{Paulson2019}. In MATLAB, this can be done easily using the commands in Algorithm~\ref{alg: tilt}.

\begin{algorithm}[h]
\caption{MATLAB code to draw random tilts}
\begin{Verbatim}[fontsize = \footnotesize]
kappa2 = I0(1)*7.7554*(D/r0)^(5/3)/(2^(5/3))*...
        (2*lambda/(pi*D))^2*(2*D/lambda)^2;
Cx     = kappa2*C2;
Cxfft  = fft2(Cx);
S_half = sqrt(abs(Cxfft));
b   = randn(N,N);
MVx = real(ifft2(S_half.*b))*sqrt(2)*N;
\end{Verbatim}
\label{alg: tilt}
\end{algorithm}

Two additional detailed issues need to be taken care of.
\begin{itemize}
\item FFT scaling. In the last line of the algorithm there is a factor \texttt{N} and \texttt{sqrt(2)}. The factor \texttt{N} accounts for the scaling of \texttt{fft2}, and \texttt{sqrt(2)} accounts for the normalization of the real part.
\item Smoothing the tilts. The tilts generated from the power spectral density are spatially correlated but could be noisy. A simple way to improve the smoothness of the tilts is to truncate the small entries of the power spectral density. The cutoff threshold for the truncation is typically 1\% of the peak of the power spectral density. Alternatively, one may extend the range of $s$ or apply a windowing function to $C_i(\vxi)$.
\end{itemize}

Thus far we have discussed how to draw random tilts from the correlation coefficients. However, throughout the development, in particular \eref{eq: correlation 3}, we have not yet talked about the range of $s$. The relationship between the range of $s$ and other system parameters is defined by the approximation in Lemma 3. Specifically, we need
\begin{equation}
Ds = L |\vtheta - \vtheta'|.
\end{equation}
The physical meaning of $L |\vtheta - \vtheta'|$ is the displacement in the object plane. Recall that the Nyquist pixel spacing in the object plane is $L \lambda / (2D)$. Thus, if we assume that the smallest displacement caused by $L |\vtheta - \vtheta'|$ is exactly the Nyquist spacing $L \lambda / (2D)$, then we can determine the per-unit spacing of $s$.
\begin{equation}
s = \frac{L|\vtheta - \vtheta'|}{D} = \frac{\frac{L\lambda}{2D}}{D} = \frac{L \lambda }{2D^2}.
\end{equation}
As a result, the maximum $s$ is given by
\begin{equation}
s_{\max} = \frac{L \lambda }{2D^2} N,
\end{equation}
where $N$ is the number of pixels of the image. For the configurations in Example 1: $\lambda = 525$nm, $L = 7000$m, $D = 0.2034$m, and $N = 512$ pixels, we have that $s_{\max} \approx 22.74$. 

\subsection{Spatial Correlations of High-Order Zernikes}
For the previous two sections we have been primarily concerned with the correlation of the first two Zernike coefficients. This is due to the fact that there exits a relatively simple expression for obtaining this correlation through Chanan \cite{chanan92}. We now turn to the correlation of the higher order coefficients, which is not as convenient to write but quite straight-forward to compute.

The extension to high order coefficients is obtained through the work of Takato et al. \cite{Takato1995}. The idea is to use an approach nearly identical to Chanan's, but using a more general expression. Following the notation of Takato et al.\cite{Takato1995}, the spatial correlations of the Zernike coefficients are \cite[Eq. 12]{Takato1995}:
\begin{align}
    &\E[a_j^* a_{j'}(D\vxi)] \label{eq: Takato}\\
    &= 2^{14/3}\pi^{8/3}A(D/2)^{5/3}[(n+1)(n'+1)]^{1/2} f_{jj'}(s,\psi_0,1/L_0) \notag
\end{align}
with $A = 0.00969 \left(\frac{2\pi}{\lambda} \right)^2 \int_0^{\infty} C_n^2(z)dz$ and $n$ corresponding to the radial degree related to coefficient $j$, and $L_0$ as the standard notation for the outer scale size related to the Kolmogorov theory of turbulence \cite{Kolmogorov1941}. The function $f_{jj'}(s,\psi_0,1/L_0)$ is rather tedious to express, and so we refer readers to \cite[Eq. 13]{Takato1995}. We note that $f_{jj'}(s,\psi_0,1/L_0)$ is a ``5-way" conditional statement dependent on the angular degree of the associated Zernike coefficient. Therefore, computationally it is not as difficult as it may seem at first glance.

With the expression provided by Takato et al. \cite{Takato1995}, it becomes straight-forward to evaluate the spatial correlations of the higher order Zernike coefficients. Specifically, we follow all the derivations in the previous subsections but replace \eref{eq: correlation 3} with \eref{eq: Takato}. By establishing this result we can now derive spatial correlations of \emph{all} Zernike coefficients, which could be a useful result beyond building a simulator.

\subsection{Selecting High-Order Zernike Block Size}
\begin{figure}[h]
\centering
\includegraphics[width=\linewidth]{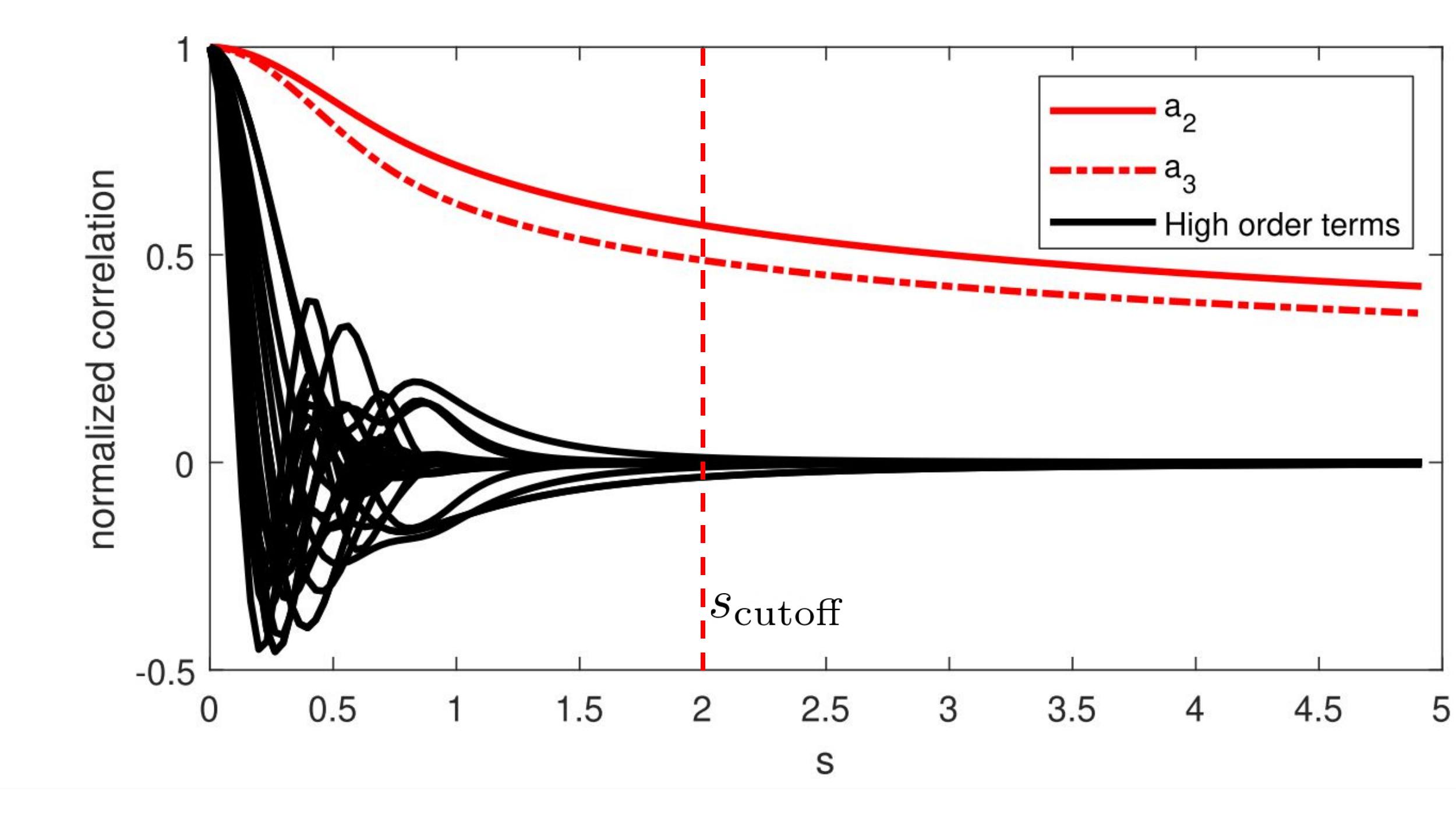}
\caption{Normalized spatial correlation of the first 36 Zernike coefficients. $a_2$ and $a_3$ represent the tilts, and the rest represents the high-order abberations. Note the strong contrast between the tilts and the high-order terms for the spread in the spatial correlation.}
\label{fig: Takato}
\end{figure}

While we have just showed that it is \emph{possible} to impose spatial correlations to all Zernike coefficients, in practice we observe that it is enough to just focus on the correlations of the first two coefficients. To illustrate this idea we plot the normalized spatial correlations as a function of the proportion factor $s$ in \fref{fig: Takato}. It is quite surprising to see the strong contrast between the tilts ($a_2$ and $a_3$) and the other high-order aberration terms $a_4, \ldots, a_{36}$. The tilts have a steady level of spatial correlation even for very large $s$, whereas the high-order aberration terms decay very quickly. The implication is that high-order aberration terms do not have a long-range correlation, and so we can approximate them with independent samples over blocks of the image.

If we draw independent high-order Zernike coefficients for different blocks of pixels in the image, how should we choose the block size? The number of blocks is determined by a cutoff $s_{\text{cutoff}}$ as shown in \fref{fig: Takato} ($s_{\text{cutoff}} = 2$ has approximately vanishing correlation). To understand how $s_{\text{cutoff}}$ interacts with the image, we first note that $s$ is a quantity exclusively for the geometry and not the turbulence level. Readers familiar with the split-step approach can observe that the correlation between two sources is established based on the amount of overlaps in the phase screens across the propagation path. In our framework this can be measured in terms of the value of $s$, with small $s$ corresponding to a large overlap, and large $s$ with small overlap. Of course, there is always \emph{some} overlap near the aperture, and so the correlations need not vanish. Analogously, $s$ does not go to infinity as we move across the FOV.

Given a value $s_{\text{cutoff}}$, we can ask about its relationship with the isoplanatic angle $\theta_0$. If $\theta_0 \geq s_\text{cutoff}$, i.e., the block is smaller than the isoplanatic angle, then drawing independent high-order Zernike coefficients is largely valid \cite{HardieSimulator}. If $\theta_0 < s_\text{cutoff}$, i.e., the block is larger than the isoplanatic angle, then our independent sampling will cause inaccuracy. We acknowledge this limitation as a trade-off between accuracy and speed. Should this becomes a necessary problem, we can either reduce $s_\text{cutoff}$ or we can interpolate the phase before forming the PSF.

\begin{figure}[h]
\centering
\includegraphics[width=\linewidth, trim=1cm 0 1cm 0, clip]{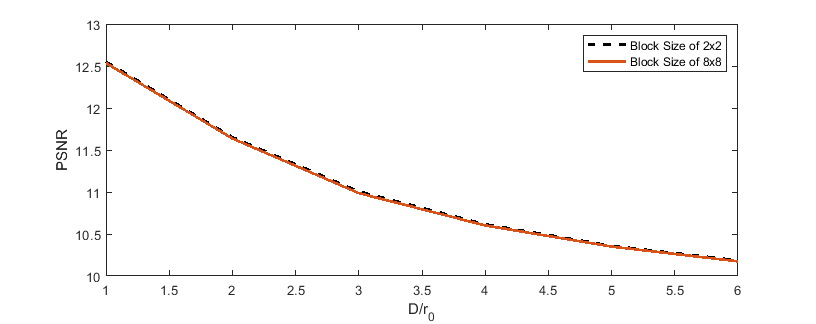}
\caption{PSNR between ground truth image and images formed using block sizes of $2 \times 2$ and $8 \times 8$. For varying levels of turbulence on a test pattern, we can see through PSNR the cases are indistinguishable.}
\label{fig: PSNR_blocks}
\end{figure}

From our experience, $s_\text{cutoff}$ usually corresponds to a small amount of pixels when compared to the entire image. Additionally, the domain where $\theta_0 < s_\text{cutoff}$ would typically correspond to medium to high turbulence. Even if the coefficients are correlated, their summed behavior is highly varying enough that it would be difficult to distinguish from again partitioning the block size and assuming independence. From an image quality perspective, partitioning the blocks smaller than $s_\text{cutoff}$ does not have any observable difference. In \fref{fig: PSNR_blocks} we show the peak signal to noise ratio between a clean image and a turbulent image simulated according to two different block sizes $2 \times 2$ and $8 \times 8$. It can be observed that at all the typical turbulence levels $D/r_0$, the PSNR values of the two choices of block sizes are almost identical. This suggests that in the pixel domain the influence of $s_\text{cutoff}$ is insignificant for most practical situations.

\subsection{Overall Simulator}
The overall simulator can be summarized in the following steps. We first partition the image into blocks and draw the point spread functions using Algorithms~\ref{alg: phase screen} and \ref{alg: random sample}. We then apply spatially varying blur to the blocks. Afterwards, we draw the tilts using Algorithm~\ref{alg: tilt} (this requires setting the covariance matrix using Algorithms~\ref{alg: I0I2} and Algorithm~\ref{alg: CxCy}.) We then warp the image according to the random motion vectors. The complete MATLAB code is available at [TBD after peer-review.]

\section{Validations}
We report experimental results in this section. Our approach is to conduct a quantitative analysis by comparing what the theory predicts and what the simulator generates. We argue that this is the only way to verify the correctness of the simulator in the absence of ground truth. We will show visual comparisons with field data, but these should only be considered as side evidence. We should also emphasize that many of the ``ground truth'' datasets in the literature cannot be used because they are generated using a hot-air burner \cite{Anantrasirichai2013, Milanfar2013}, and do not follow the long-range statistics.

\begin{figure*}[!]
\centering
\begin{tabular}{cc}
\includegraphics[width=0.47\linewidth]{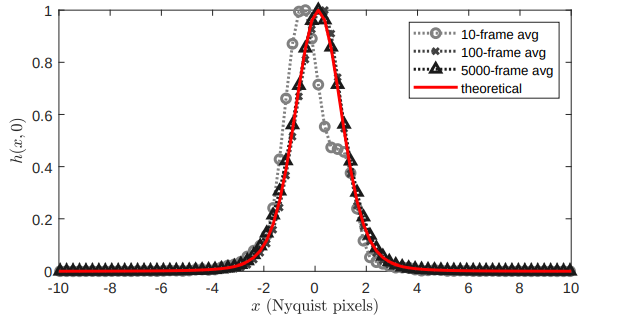}&
\includegraphics[width=0.47\linewidth]{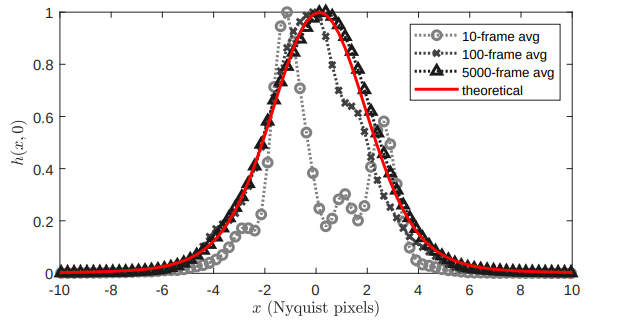}\\
Long-exposure, $C_n^2 = 2.5\times10^{-16}$m$^{-2/3}$ & Long-exposure, $C_n^2 = 1\times10^{-15}$m$^{-2/3}$\\
\includegraphics[width=0.47\linewidth]{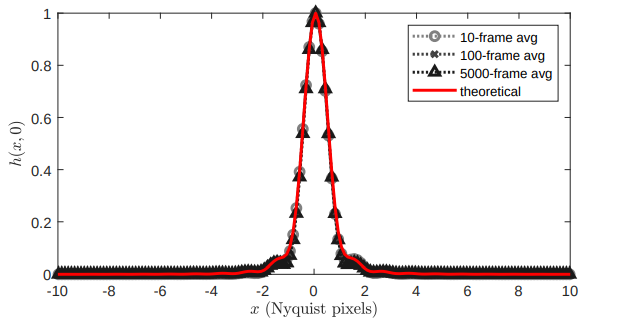}&
\includegraphics[width=0.47\linewidth]{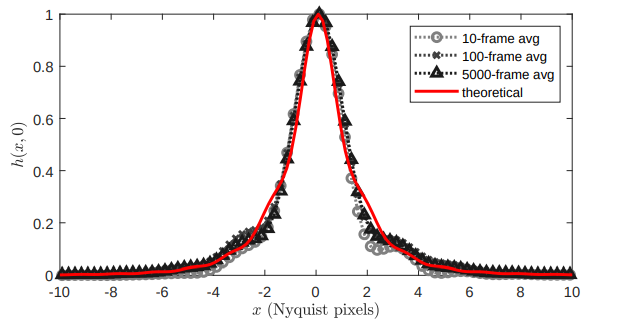}\\
Short-exposure, $C_n^2 = 2.5\times10^{-16}$m$^{-2/3}$ & Short-exposure, $C_n^2 = 1\times10^{-15}$m$^{-2/3}$
\end{tabular}
\caption{The theoretical long-exposure PSF and short exposure PSF compared with the empirical averaged instantaneous PSFs for $C_n^2 = 2.5\times10^{-16}$m$^{-2/3}$ and $C_n^2 = 1\times10^{-15}$m$^{-2/3}$. Shown in each sub-figure are the 10-frame average, 100-frame average and 5000-frame average PSFs, all in black. The red curve is the theoretically predicted result. The $x$-axis denotes the Nyquist pixel, where one pixel correspond to $\lambda d/2D$. Note the excellent match between the theoretically predicted curves and the simulated curves. }
\label{fig: Long Exposure 2}
\vspace{-2ex}
\end{figure*}

\subsection{Experimental Setup}
The optical parameters of our experiments are listed in Table~\ref{table: parameters}. These configurations are similar to Hardie et al. \cite{HardieSimulator}. In particular, we are interested in Kolmogorov structure function where the outer-scale of the turbulence is large. The range of $C_n^2$ is from $1\times 10^{-16}$m$^{-2/3}$ to $2.5\times 10^{-15}$m$^{-2/3}$, typical for weak to moderate turbulence levels. We consider the averaged $C_n^2$'s over the path length. The corresponding theoretical values of the Fried parameter $r_0$ and the isoplanatic angle $\theta_0$ are listed in the bottom of the Table.

\begin{table}[h]
\caption{Simulation Parameters}
\label{table: parameters}
\centering
\begin{tabular}{ll}
\hline\hline
Parameter & Value \\
\hline
Path length         & $L$ = 7km\\
Aperture Diameter   & $D$ = 0.2034m\\
Focal Length        & $d$ = 1.2m\\
Wavelength          & $\lambda$ = 525nm\\
Zernike Phase Size  & $64 \times 64$ pixels\\
Image Size          & $512 \times 512$ pixels\\
Nyquist spacing (object plane) $\frac{L\lambda}{2D}$ & $\delta_0 = 9.0344$mm\\
Nyquist spacing (focal plane) $\frac{d\lambda}{2D}$ & $\delta_f = 1.5488\mu$m\\
\end{tabular}

\vspace{1ex}
\small{
\begin{tabular}{cccccc}
\hline\hline
 & \multicolumn{5}{c}{$C_n^2$ $\times10^{-15}$ (m$^{-2/3}$)}\\
 & $0.1$ & $0.25$ & $0.5$ & $1.0$ & $2.5$\\
\hline
$r_0$ (m)       & 0.1901 & 0.1097 & 0.0724 & 0.0478 & 0.0374\\
$\theta_0$ ($\mu$rad)  & 8.5401 & 4.9283 & 3.2515 & 2.1452 & 1.6819\\
$\theta_0$ (pixel)     & 6.6170 & 3.8186 & 2.5193 & 1.6621 & 1.3032\\
\hline
\end{tabular}}
\vspace{-2ex}
\end{table}

\subsection{Theoretical Long and Short Exposures}
We first compare the empirical averaged PSFs with the ideal long-exposure PSF. We generate $H_{\mathrm{atm}}^{\phi}$ according to the tilt and the high-order Zernike coefficients, and then consider the diffraction limited blur to construct $H(\vf) = H_{\mathrm{dif}}(\vf)H_{\mathrm{atm}}^{\phi}(\vf)$. The instantaneous PSF is the inverse Fourier transform of $H(\vf)$. The process is repeated 5000 times to generate 5000 independent PSFs. Empirical average is then compared with the theoretical long-exposure.

\begin{figure}[t]
\vspace{-2ex}
\centering
\footnotesize{
\begin{tabular}{cccc}
\includegraphics[width=0.23\linewidth]{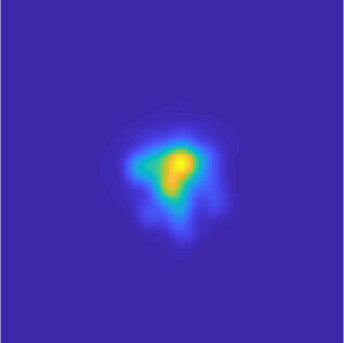}&
\hspace{-2ex}\includegraphics[width=0.23\linewidth]{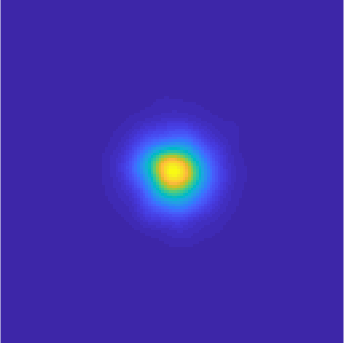}&
\hspace{-2ex}\includegraphics[width=0.23\linewidth]{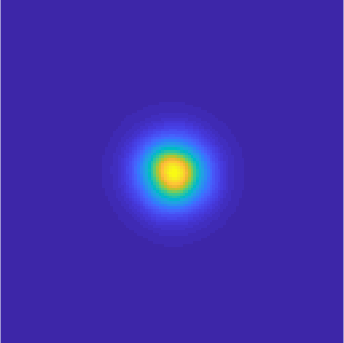}&
\hspace{-2ex}\includegraphics[width=0.23\linewidth]{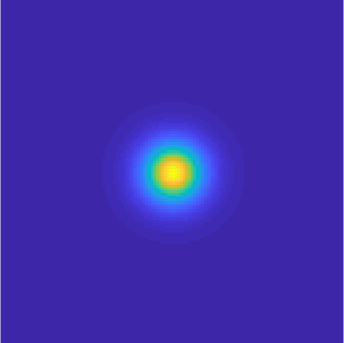}\\
(a) 50-avg &(b) 500-avg &(c) 5000-avg &(d) Theoretical\\
\end{tabular}}
\caption{Instantaneous PSFs generated using the proposed simulator for $C_n^2 = 1\times 10^{-15}$m$^{-2/3}$. (a) 50-frame average. (b) 500-frame average. (c) 5000-frame average. (d) Theoretical long-exposure PSF.}
\vspace{-2ex}
\label{fig: Long Exposure}
\end{figure}

\fref{fig: Long Exposure 2}(a)-(b) shows the cross sections of the empirical average of the instantaneous PSFs and the theoretical long-exposure PSF. As the number of frames increases, the empirical average approaches to the theoretical PSF quickly. The exact match of the two at asymptotic limit suggests that the simulated results are consistent with the theory.

The same experiment is repeated for short-exposure PSFs. In this case, we only draw high-order coefficients. This leads to $H(\vf) = H_{\mathrm{dif}}(\vf)H_{\mathrm{atm}}^{\varphi}(\vf)$, where $H_{\mathrm{atm}}^{\varphi}(\vf)$ is the atmospheric OTF without the tilt. The theoretical short-exposure OTF follows \eref{eq: HSE}. The results of this experiment are shown in \fref{fig: Long Exposure 2}(c)-(d). As expected, the short-exposure PSF is narrower than the long-exposure PSF. However, the empirical average of the instantaneous PSFs still converge quickly to the theoretical short-exposure PSF, indicating the match between the theory and the algorithm.

\fref{fig: Long Exposure} shows the case where $C_n^2 = 1\times 10^{-15}$m$^{-2/3}$. The sub-figures correspond to 50-frame average, 500-frame average, and 5000-frame average. Note the quickly converging instantaneous PSFs to the theoretical long-exposure PSF.

\begin{figure*}[!]
\centering
\footnotesize{
\begin{tabular}{cc}
\includegraphics[width=0.475\linewidth]{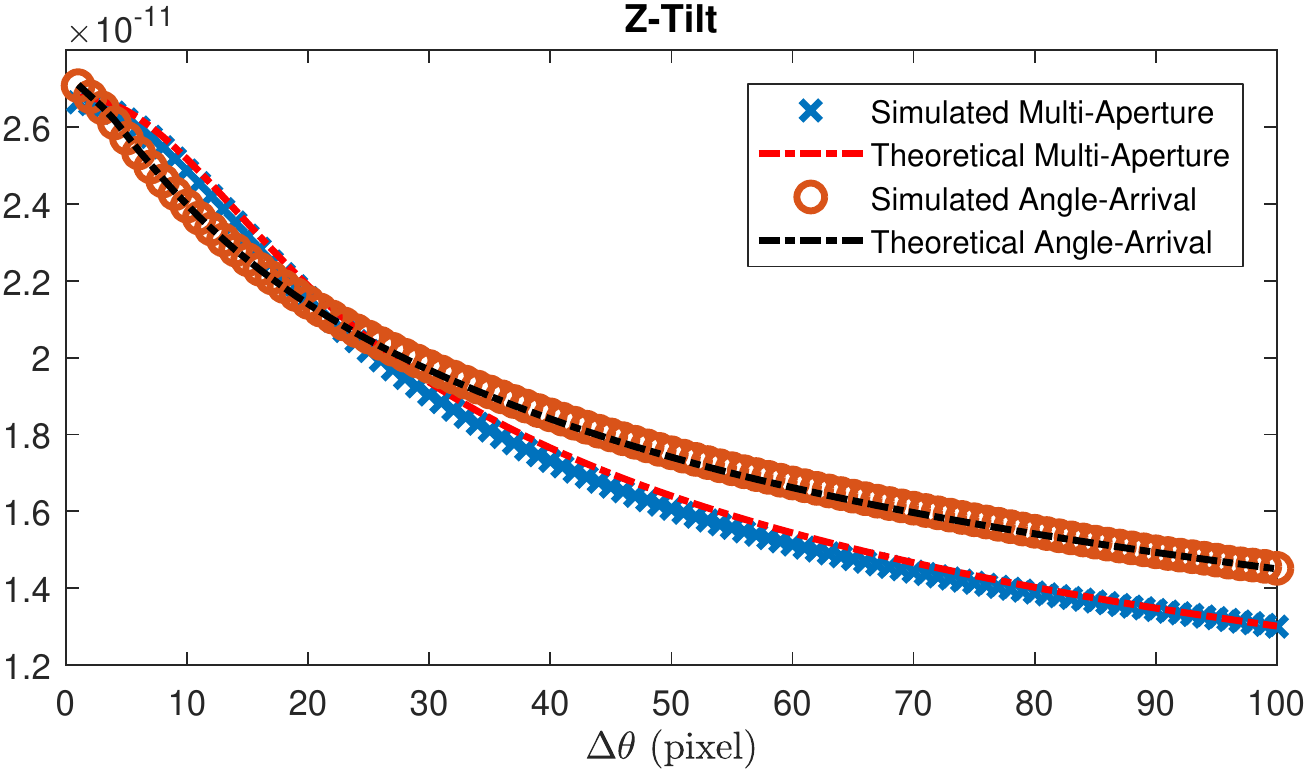}&
\includegraphics[width=0.475\linewidth]{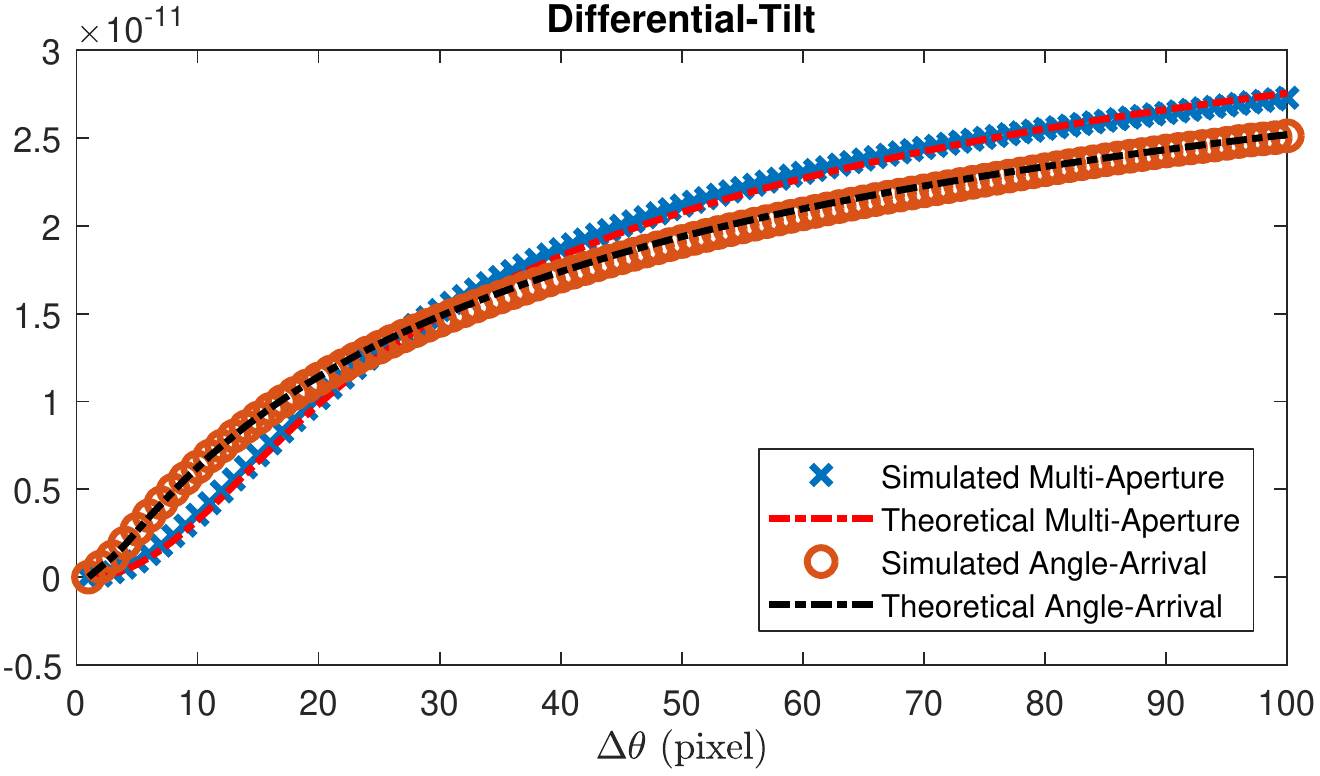}
\end{tabular}}
\caption{The theoretical and simulated tilt statistics at $C_n^2 = 1 \times 10^{-15}$m$^{-2/3}$. The curves are plotted versus the number of Nyquist pixels. The angle-of-arrival is due to Basu et al. \cite{Basu_2015}, and the multi-aperture is our approximation based on Chanan \cite{chanan92}. Additionally, we note $20,000$ random realizations were used in generating the simulated curves.}
\label{fig: Ztilt and DTV}
\end{figure*}

\begin{figure*}[!]
\centering
\begin{tabular}{cccc}
\hspace{-2ex}\includegraphics[width=0.25\linewidth]{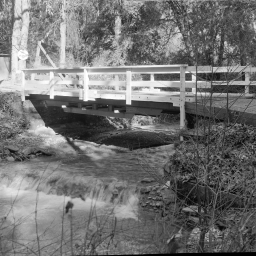}&
\hspace{-2ex}\includegraphics[width=0.25\linewidth]{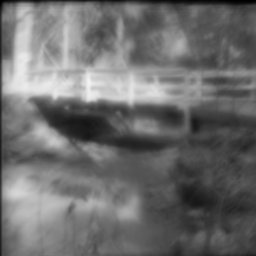}&
\hspace{-2ex}\includegraphics[width=0.25\linewidth]{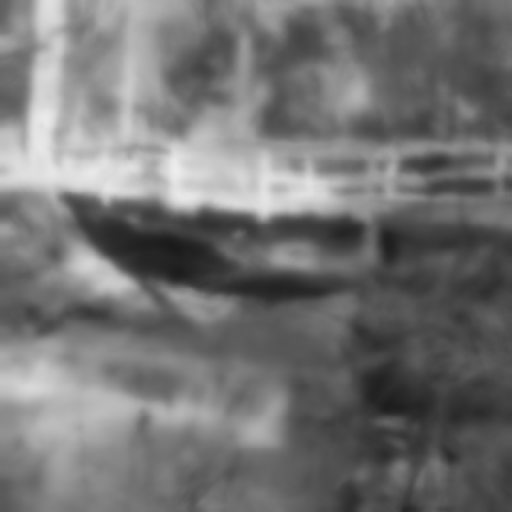}&
\hspace{-2ex}\includegraphics[width=0.25\linewidth]{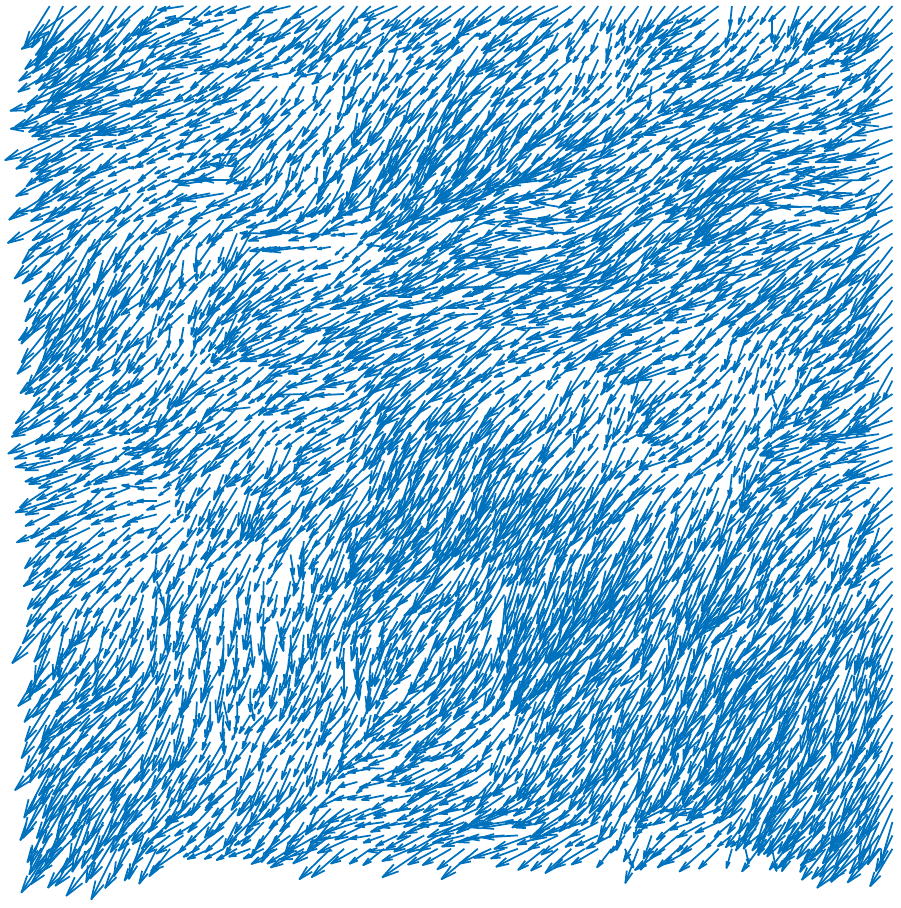}\\
\hspace{-2ex}\includegraphics[width=0.25\linewidth]{./pix/bridge}&
\hspace{-2ex}\includegraphics[width=0.25\linewidth]{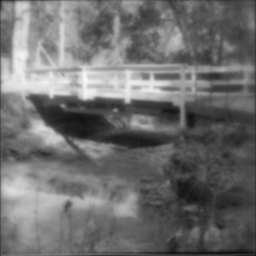}&
\hspace{-2ex}\includegraphics[width=0.25\linewidth]{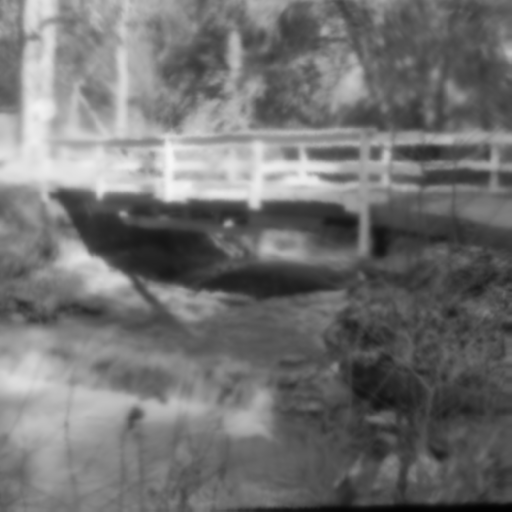}&
\hspace{-2ex}\includegraphics[width=0.25\linewidth]{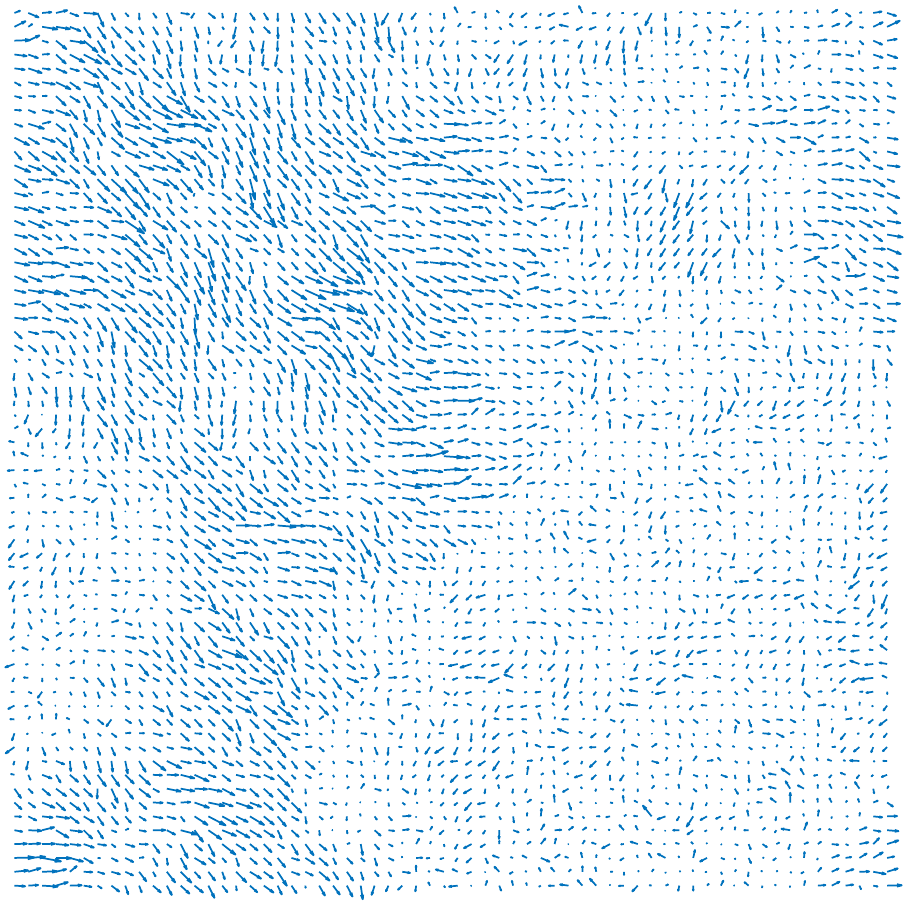}\\
(a) Clean &(b) Split-Step \cite{HardieSimulator} &(c) Ours &(d) Our tilt map\\
\end{tabular}
\caption{Simulated images and tilt maps for [Top] $C_n^2 = 1\times 10^{-15}$m$^{-2/3}$, and [Bottom] $C_n^2 = 2.5\times 10^{-16}$m$^{-2/3}$. The tilts are scaled by $2\times $ and skipped per every 8 pixels for display.}
\label{fig: Bridge}
\end{figure*}

\subsection{Theoretical Tilt Statistics}
The second experiment is to verify the Z-tilt correlation and the differential tilt variance (DTV), two statistical quantities that can provide theoretical ground truth statistics for comparison. The Z-tilt correlation measures how the tilts are correlated as the angle-of-arrival increases. DTV measures the variance of the tilts as a function of the angle-of-arrivals. The definitions of these two quantities can be found in \cite[Eq. 8]{HardieSimulator} and \cite[Eq. 14]{HardieSimulator}.

We fix $C_n^2 = 1 \times 10^{-15}$m$^{-2/3}$ and vary the angle-of-arrival $\Delta \theta$ from 0 pixel to 100 pixels. The results are shown in \fref{fig: Ztilt and DTV}, where we plot both the tilts drawn from the angle-of-arrival statistics by Basu et al. \cite{Basu_2015} and the multi-aperture statistics of this paper. Despite the small discrepancies due to the approximation in Lemma 3, the shapes and trends of the correlations match very well, providing further justification of the proposed simulator.

\subsection{Comparison with Split-Step Propagation \cite{HardieSimulator}}
This experiment provides a visual comparison with the split-step method reported by Hardie et al. \cite{HardieSimulator}. The experiment involves simulating turbulence for two different $C_n^2 = 1\times10^{-15}$m$^{-2/3}$ and $2.5\times10^{-16}$m$^{-2/3}$. The image we use for testing is the \texttt{bridge} image of size $512 \times 512$. Other testing configurations follow Table~\ref{table: parameters}.

The results are shown in \fref{fig: Bridge}, where we show the clean image, the simulated turbulence, and the corresponding tilt map. For visualization we scale the tilts by $2\times$, and skip the tilts for every 8 pixels. Readers are encouraged to compare with Fig. 16 of Hardie et al. \cite{HardieSimulator}. In brief, the tilts and blurs generated by the proposed simulator are visually very similar to those generated by Hardie et al. \cite{HardieSimulator}. We would like to additionally note that the images shown do not incorporate differences in length of propagation across the image.


\begin{figure*}[!]
\centering
\begin{tabular}{ccc}
\hspace{-2ex}\includegraphics[width=0.33\linewidth]{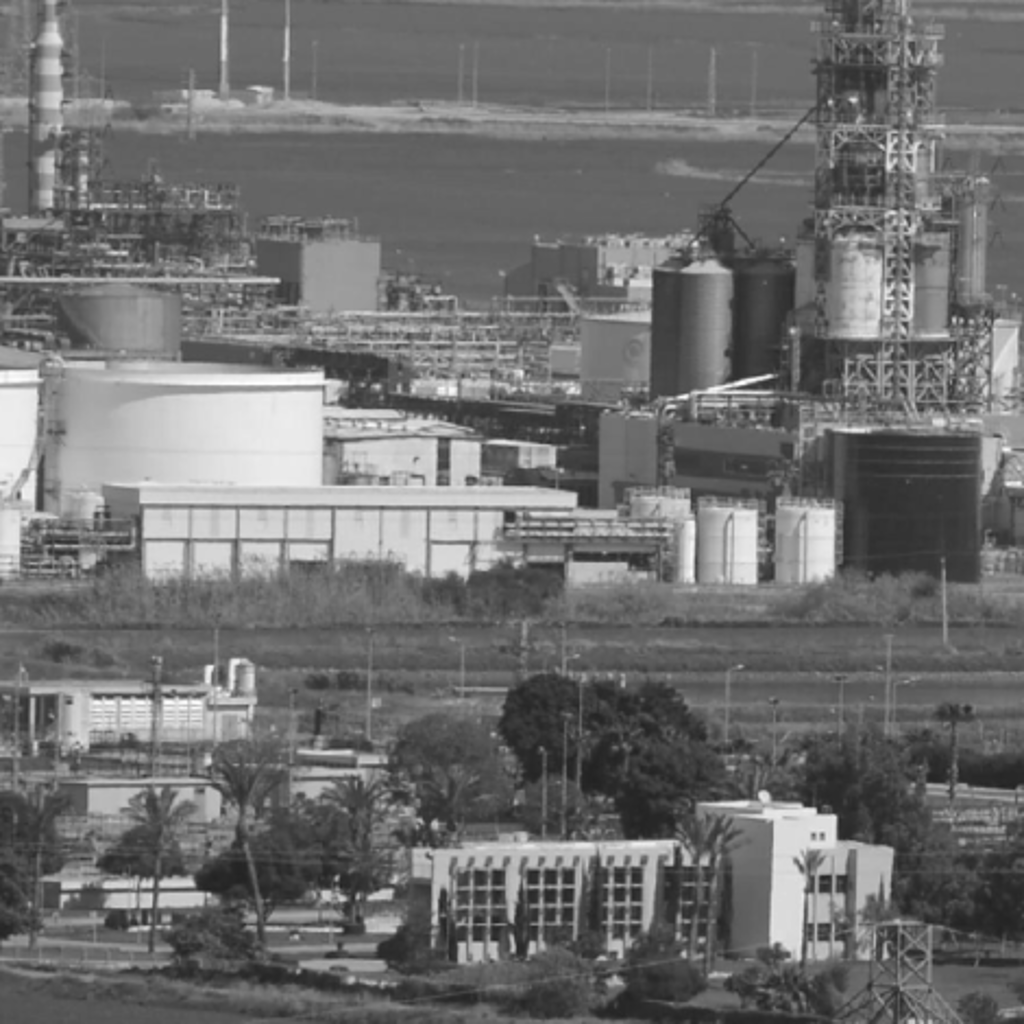}&
\hspace{-2ex}\includegraphics[width=0.33\linewidth]{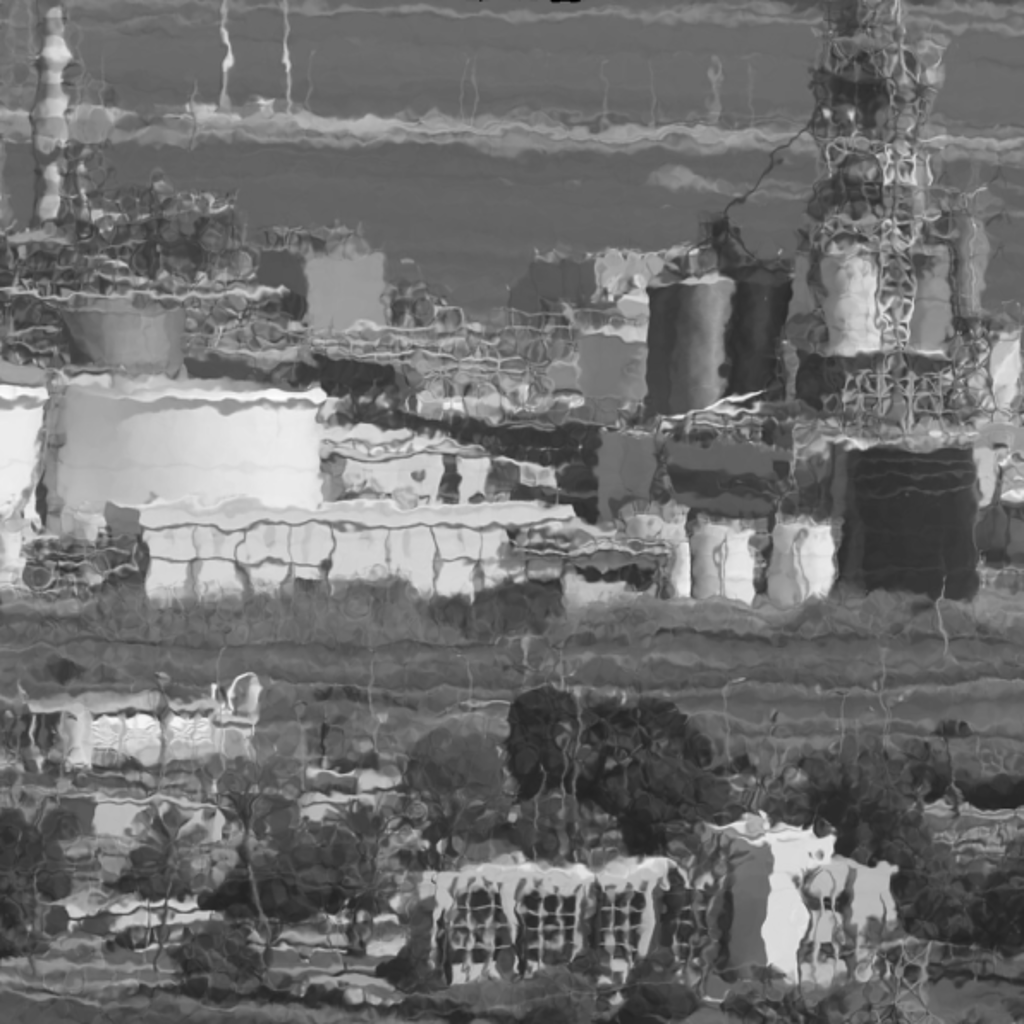}&
\hspace{-2ex}\includegraphics[width=0.33\linewidth]{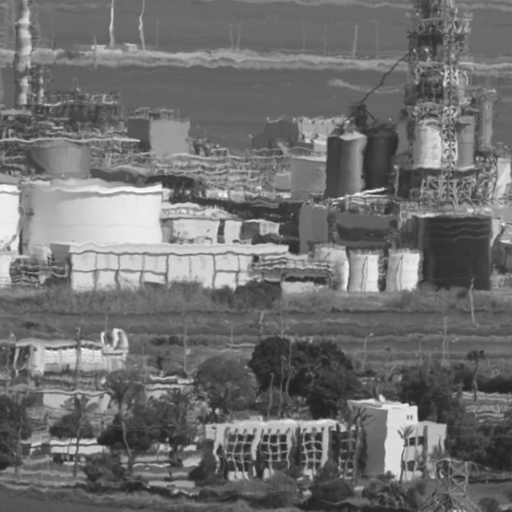}\\
(a) Clean input &(b) \cite[Fig. 10]{Schwarzman2017ICCP} &(c) Ours, tilt-only
\end{tabular}
\caption{Comparison with Schwartzman \cite{Schwarzman2017ICCP}. The optical parameters are $C_n^2 = 3.6 \times 10^{-13}$m$^{-2/3}$, $L = 2000$m, $d = 0.3$m, and $D = 0.054$m. The pixel spacing is $8\times$ the Nyquist pixel spacing.}
\label{fig: Plant}
\end{figure*}

\subsection{Comparison with Schwartzman et al. \cite{Schwarzman2017ICCP}}
Another visual comparison we make is with Schwartzman et al. \cite{Schwarzman2017ICCP}. This simulation uses $C_n^2 = 3.6 \times 10^{-13}$m$^{-2/3}$, with $L = 2000$m. The focal length of the lens is $d = 0.3$m, and the aperture diameter is $D = 0.054$m. The testing image size is $512 \times 512$. In simulating this image, we scale the Nyquist pixel spacing by $8 \times$ to reflect the field of view of the scene. Schwartzman et al. did not provide this spacing information, but we think that the spacing is critical for otherwise the warping will be excessive or insufficient.

The results of this experiment are shown in \fref{fig: Plant}. \fref{fig: Plant}(a) shows the ground truth clean image, and \fref{fig: Plant}(b) shows the simulated result of \cite{Schwarzman2017ICCP}. \fref{fig: Plant}(c) shows the tilt-only simulation using our proposed simulator. Visually, we observe that the two results are similar to some extent. However, we should note that the simulator by Schwartzman et al. does not simulate spatially varying blur whereas our simulator does.

\subsection{Comparison with Real Data}
We compare our simulated images with real turbulence images. We use the data reported in \cite{Leonard_Howe_Oxford}. The images were taken during a North Atlantic Treaty Organization (NATO) field trial at White Sands Missile Range in New Mexico, USA in 2005. The optical parameters follow the descriptions in \cite[Fig. 1]{Leonard_Howe_Oxford}, where the focal length is $d = 0.8$m and the aperture diameter is $D = 0.05$m. We assume the wavelength is $\lambda = 525$nm, which is approximately in the middle of white light. The target is located at $L = 1000$m from the camera. The $C_n^2$ values range from $2\times 10^{-14}$m$^{-2/3}$ to $8 \times 10^{-13}$m $^{-2/3}$, corresponding to sunrise and noon of the day. For simulation, we used \cite[Fig. 4a]{Leonard_Howe_Oxford} as the ``ground truth'', and simulate the turbulence effects for higher turbulence levels. The results are compared with the field data.



\begin{figure*}[t]
\centering
\includegraphics[width=0.95\linewidth]{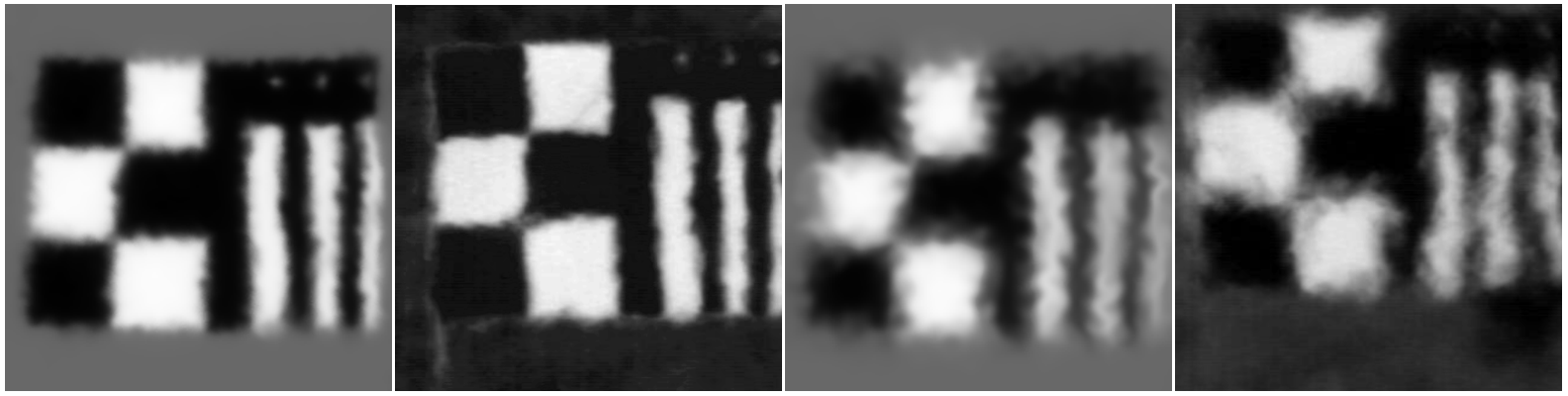}
\caption{NATO RTO SET/RTG-40 dataset reported by Leonard, Howe and Oxford in \cite{Leonard_Howe_Oxford}. The optical parameters used in this simulation are $D = 0.05$m, $d = 0.8$m, $\lambda = 525$nm, and $L = 1000$m. [Left Two] Simulated image and field data for medium turbulence at $C_n^2 = 1\times10^{-13}$ m $^{-2/3}$. [Right Two] Simulated image and field data for strong turbulence at $C_n^2 = 8\times10^{-13}$ m $^{-2/3}$. We also note we have contrast corrected the field data images for ease of visual comparison.}
\label{fig: Pattern1}
\end{figure*}

The results of this experiment are reported in \fref{fig: Pattern1}, where we show two cases of $C_n^2 = 1\times10^{-13}$ m $^{-2/3}$ and $C_n^2 = 8\times10^{-13}$ m $^{-2/3}$. In this simulation, we scaled the Nyquist pixel spacing by $2 \times$ to match the target size. As we can observe, the simulated turbulence has a good visual similarity compared to the field data.

\subsection{Implementation}
The simulator is implemented in MATLAB 2017b. The computer used to run the simulation is equipped with Intel(R) Core(TM) i7-4770 CPU at 3.4GHz, with 32GB RAM and has a Windows 7 OS. The runtime is listed in Table~\ref{tab: runtime}. Among the time listed, Zernike PSF generation takes as much time as the spatially varying convolution. The tilt generation and the warping are insignificant compared to the blur. The split-step method implemented by Hardie et al. \cite{HardieSimulator} has a much finer PSF grid of $64 \times 64$ for an $256\times 256$ image. In the proposed simulator, the grid size can be significantly smaller because the tilt is taken care by the tilt correlation which does not require generating the Zernike phase. Note also that the run time reported here is based on a CPU. Hardie et al. \cite{HardieSimulator} uses a GPU.

\begin{table}[h]
\centering
\caption{Average run time for processing one simulated $512 \times 512$ pixel frame.}
\label{tab: runtime}
\begin{tabular}{lcc}
\hline\hline
Component & Grid Size & Run time (s)\\
\hline
Zernike PSF generation      & $8 \times 8$      & 1.11\\
                            & $16 \times 16$    & 3.31 \\
                            & $32 \times 32$    & 11.16 \\
\hline
Tilt generation and warp    & $512 \times 512$  & 1.84 \\
\hline
Spatial variant convolution & $8 \times 8$      & 1.13 \\
                            & $16 \times 16$    & 3.10 \\
                            & $32 \times 32$    & 9.53 \\
\hline
Total (w/o GPU)             & $8 \times 8$      & 4.08\\
                            & $16 \times 16$    & 8.25\\
                            & $32 \times 32$    & 22.53\\
\hline
\end{tabular}
\end{table}

\section{Conclusion}
A new simulator for anisoplanatic turbulence is proposed. The simulator decouples the tilts and high-order abberations of the wavefront distortion using Zernike decompositions. The tilts are drawn according to a spatial covariance matrix, whereas the high-order abberations are drawn according to an inter-mode covariance matrix. By leveraging the homogeneity of the tilts, fast Fourier transforms are used to enable efficient sampling of the tilts. The number of spatially varying blurs is substantially fewer than traditional wave propagation methods, as the tilts have been handled separately. Experimental results have confirmed the efficiency and accuracy of the simulator. It is anticipated that the new simulator can significantly improve our ability to evaluate turbulence mitigation algorithms, and improve our ability to generate training samples for learning-based algorithms.

\subsection*{Acknowledgement}
A shorter version of the paper has been presented in the IEEE International Conference on Computational Photography, 2020 \cite{Chimitt_ICCP_2020}.

The work is funded, in part, by the Air Force Research Lab and Leidos, and by the National
Science Foundation under grants CCF-1763896 and CCF-1718007. The authors would like to thank Michael Rucci, Barry Karch, Daniel LeMaster and Edward Hovenac of the Air Force Research Lab for many insightful discussions.

This work has been cleared for public release carrying the approval number 88ABW-2019-5034.

\section*{Appendix}
\subsection*{Proof of Lemma 1}
Since $Z_1(\vrho) = 1$, it holds that $\int W(\vrho) Z_j(\vrho) d\vrho = \int W(\vrho) Z_j(\vrho) Z_1(\vrho) d\vrho = 0$, where the last equality is due to orthogonality of the Zernike polynomials. \hfill $\square$

\subsection*{Proof of Lemma 2}
Expand the structure function as
\begin{equation*}
D_{\phi}(R\vrho-R\vrho') = \E[\phi(R\vrho)^2] - \E[\phi(R\vrho)  \phi(R\vrho')] + \E[\phi(R\vrho')^2].
\end{equation*}
Substitute this into \eref{eq: correlation 2}. The double integration involving $\E[\phi(R\vrho)^2]$ can be simplified to a product of two integrals
\begin{align*}
&\int \int W(\vrho)W(\vrho') Z_j(\vrho) Z_{j'}(\vrho')\E[\phi(R\vrho)^2] d\vrho d\vrho'\\
&= \int W(\vrho) Z_j(\vrho) \E[\phi(R\vrho)^2] d\vrho \underset{= 0 }{\underbrace{\int W(\vrho')Z_j(\vrho') d\vrho'}},
\end{align*}
where the latter integral vanishes due to Lemma 1. The same argument applies to $\E[\phi(R\vrho')^2]$. This proves the lemma. \hfill $\square$

\subsection*{Proof of Lemma 3}
Consider the integral in \eref{eq: D new}:
$$
I = \int_0^L  \left| (R\vrho-R\vrho')\left(1-\frac{z}{L}\right) + z(\vtheta-\vtheta')\right|^{5/3} dz.
$$
Let $u = \frac{z}{L}$, and define
$$f(u) = \left| (R\vrho-R\vrho')\left(1-u\right) + Lu(\vtheta-\vtheta')\right|^{5/3}.$$
The first order Taylor approximation of $f(u)$ at $u = 1/2$ is
\begin{equation*}
f(u) \cong f\left(\frac{1}{2}\right) + f'\left(\frac{1}{2}\right)\left(u-\frac{1}{2}\right).
\end{equation*}
Integrating $f(u)$ from $0$ to $1$, and recognizing that $\int_0^1 (u-1/2) du = 0$, we have
\begin{align*}
I = L \int_{0}^1 f(u) du = L \left| \frac{(R\vrho-R\vrho')}{2} + \frac{L (\vtheta-\vtheta')}{2} \right|^{5/3}.
\end{align*}
Substituting into \eref{eq: D new} yields \eref{eq: D new 2}. \hfill $\square$

\bibliographystyle{IEEEbib}
\bibliography{egbib}
\end{document}